# Stellar Imaging Coronagraph and Exoplanet Coronal Spectrometer – Two Additional Instruments for Exoplanet Exploration Onboard The WSO-UV 1.7 Meter Orbital Telescope

**Alexander Tavrov[a, b], Shingo Kameda[c], Andrey Yudaev[b], Ilia Dzyuban[a], Alexander Kiselev[a], Inna Shashkova[a]*, Oleg Korablev[a], Mikhail Sachkov[d], Jun Nishikawa[e], Motohide Tamura[e, g], Go Murakami[f], Keigo Enya[f], Masahiro Ikoma[g], Norio Narita[g]**

[a] IKI-RAS Space Research Institute of Russian Academy of Science, Profsoyuznaya ul. 84/32, Moscow, 117997, Russia

[b] Moscow Institute of Physics and Technology, 9 Institutskiy per., Dolgoprudny, Moscow Region 141700, Russia

[c] Rikkyo University, 3-34-1 Nishi-Ikebukuro, Toshima, Tokyo, 171-8501, Japan

[d] INASAN Institute of Astronomy of the Russian Academy of Sciences, Pyatnitskaya str., 48 , Moscow, 119017, Russia

[e] NAOJ National Astronomical Observatory of Japan, 2-21-1 Osawa, Mitaka, Tokyo 181-8588, Japan

[f] JAXA Japan Aerospace Exploration Agency, 3-3-1 Yoshinodai, Chuo, Sagamihara, Kanagawa, 229-8510, Japan

[g] The University of Tokyo, 7-3-1 Hongo, Bunkyo, Tokyo, 113-8654, Japan

**Abstract.** The World Space Observatory for Ultraviolet (WSO-UV) is an orbital optical telescope with a 1.7 m-diameter primary mirror currently under development. The WSO-UV is aimed to operate in the 115–310 nm UV spectral range. Its two major science instruments are UV spectrographs and a UV imaging field camera with filter wheels. The WSO-UV project is currently in the implementation phase, with a tentative launch date in 2023. As designed, the telescope field of view (FoV) in the focal plane is not fully occupied by instruments. Recently, two additional instruments devoted to exoplanets have been proposed for WSO-UV, which are the focus of this paper. UVSPEX, a UV-Spectrograph for Exoplanets, aims to determine atomic hydrogen and oxygen abundance in the exospheres of terrestrial exoplanets. The spectral range is 115–130 nm which enables simultaneous measurement of hydrogen and oxygen emission intensities during an exoplanet transit. Study of exosphere transit photometric curves can help differentiate among different types of rocky planets. The exospheric temperature of an Earth-like planet is much higher than that of a Venus-like planet, because of the low mixing ratio of the dominant coolant ($CO_2$) in the upper atmosphere of the former, which causes a large difference in transit depth at the oxygen emission line. Thus, whether the terrestrial exoplanet is Earth-like, Venus-like, or other can be determined. SCEDI, a Stellar Coronagraph for Exoplanet Direct Imaging is aimed to directly detect the starlight reflected from exoplanets orbiting their parent stars or from the stellar vicinity including circumstellar discs, dust, and clumps. SCEDI will create an achromatic (optimized to 420–700 nm wavelength range), high-contrast stellocentric coronagraphic image of a circumstellar vicinity. The two instruments: UVSPEX and SCEDI, share common power and control modules. The present communication outlines the science goals of both proposed instruments and explains some of their engineering features.

* Inna Shashkova, E-mail: shi@iki.rssi.ru

## 1 Introduction

The World Space Observatory for Ultraviolet (WSO-UV) is a space observatory which includes a 1.7 m telescope equipped with instrumentation for spectroscopy and for imaging in the UV spectral range. The WSO-UV has a tentative launch date of 2023. The WSO-UV telescope can operate at visible wavelengths, thus allowing diffraction-limited imaging. Before the Large UV/Optical/IR Surveyor (LUVOIR) [1] and after the Hubble Space Telescope (HST), it will be the next largest mirror telescope for UV astronomy. Currently, the HST is the only space telescope with UV spectroscopy capability, but there is a strong possibility that HST will be retired by mid-2020 owing to other space missions such as the James Webb Space Telescope (JWST) and the Wide-Field Infrared Survey Telescope (WFIRST). A future flagship space mission with UV spectroscopy includes the LUVOIR, which is under discussion as a mission concept study. LUVOIR will have much higher capability than WSO-UV thanks to its Ø 10 m-class aperture size. LUVOIR will be able to address similar cases as WSO-UV, but it is unlikely that LUVOIR will be launched before 2035, even if LUVOIR is selected as a top priority mission for the next Decadal Survey. Thus, there will be few, if any, opportunities for UV spectroscopy in the mid to late 2020s, even after a number of detections of nearby Earth-like transiting exoplanets have been made by the Transiting Exoplanet Survey Satellite (TESS) and other ground-based surveys. WSO-UV is thus an important mission which will fill a gap in the capability of space-based UV spectroscopy in the mid to late 2020s, and will be able to provide the most interesting targets for further follow-up observations once LUVOIR is operational.

To date, more than 3,500 confirmed exoplanets have been registered [2, 3]. Most of these were discovered by indirect detection via stellar light periodic variability “transits” or via radial velocity (RV) measurements. Such RV measurements are quite similar to those used for asteroseismology [4]. Other techniques to detect exoplanets such as “coronagraphic imaging,” “microlensing,” and “timing” continue to contribute minor numbers regarding exoplanet statistics.

The transit method enables detection of stellar photons passing through the rim of an exoplanet disk and therefore characterizes the atmosphere of an exoplanet. However, the overall ratio of transiting to non-transiting exoplanets is known to be approximately 1:100, or even less for telluric planets. The direct observation of an exoplanet under sufficient elongation from its host star becomes possible by coronagraphic imaging. This method has unique potential for studying non-transiting planets optically and directly. A coronagraph associated with a two-meter-class space-based telescope enables observations with diffraction-limited resolution above the turbulence of the Earth's atmosphere. Thus, the coronagraphic technique will allow study of non-transiting exoplanets, and hopefully, future telluric Earth-like planets in the habitable zone.

Stellar coronagraph development was planned in NASA's TPF (Terrestrial Planet Finder) program, which has now split into three projects: TPF-C dedicated to coronagraphs, TPF-I dedicated to a multi-aperture interferometer, and TPF-O with an external occulter [5]. Several coronagraphic schemes have been elaborated, and many of them have been proposed for space observatories. In particular, JWST [6] and WFIRST-AFTA [7] will be equipped with modern stellar coronagraph instruments. Several engineering runs to test the coronagraph in space have been conducted by Picture-A and Picture–B [8] experiments on sounding rockets. While the rockets follow ballistic trajectories, sufficient operational time is available to test the coronagraph subsystems [9]. A follow-up Picture-C telescope is planned for a stratospheric balloon [10]. All these projects are completing important steps toward high-contrast imaging in the stellar vicinity.

The stellar coronagraph, SCEDI, was proposed as one of two additional simple instruments onboard the WSO-UV (which is equipped with a UV polished telescope mirror of 1.7 m diameter). SCEDI's goal is to test the coronagraph as an in-flight demonstrator, to photometrically characterize several known giant planets detected by the radial velocity technique, and in the best case, to image exoplanets and circumstellar discs by searching several tens of nearby stars. Coronagraph design is proposed based on the common-path achromatic rotational shear interfero-coronagraph [11, 12] which introduces a centrostellar occultation effect under a variable angle for two-image superposition (rotational shear). Rotational shear is planned to be fixed between 5 and 10 degrees; therefore, the interfero-coronagraph will have low sensitivity to static low-order aberrations from a telescope's images.

Spectrograph UVSPEX is the other additional simple instrument proposed onboard the WSO-UV. Its goal is to differentiate between different types of rocky planets. Exosphere transit photometric curves in the spectral UV range of 115–135 nm can determine the abundance of hot neutral atoms of oxygen in exospheres. The exospheres extend to more than ten times a planet's radius. Because of this extension, exospheres are detectable at stellar distances [13].

**This communication is structured as follows:** In Sect. **2**, we describe the WSO-UV mission in general. In Sect. **3**, we describe the science goals of the coronagraph instrument SCEDI and the UV spectrograph instrument UVSPEX spectrograph instrument. In Sects. **4** and **5**, we describe the technical aspects and implementation of SCEDI and UVSPEX, respectively. In Sect. **6**, we include short miscellaneous notes on how these two different instruments are operated by a single electronic unit. In **Appendix 1**, we list the fields of view (FoV) of WSO-UV's telescope and demonstrate a solution on how to compensate for large optical aberrations in the off-axial FoV up to the level of residual aberrations in the axial FoV of a Richie-Chrétien telescope.

## 2 The WSO-UV Mission

WSO-UV is a space observatory at geosynchronous orbit with inclination of approximately 40°, which includes a 1.7 m telescope equipped with instrumentation capable of carrying out spectroscopy and direct sky imaging in the UV spectral range (115–310 nm). The nominal mission lifetime is five years with an expected extension up to ten years. As the WSO-UV telescope is mainly designed for the UV spectral range, it operates at visible wavelengths, thus allowing diffraction-limited imaging. The telescope is a Ritchey–Chrétien reflective optical design with a 17 m focal length to access a field of view of 30 arcmin [14, 15].

WSO-UV will use the Russian NAVIGATOR platform from the Lavochkin Science & Technology Association. The platform is a unified module intended for several missions including the Radioastron (Spectrum-R, launched in 2011), the Spektrum–Roentgen–Gamma (the launch is scheduled for 2018) and the WSO-UV. The NAVIGATOR platform weighs 1,300 kg and allows for a payload mass of 1,600 kg. With its fine guidance system, the pointing stabilization accuracy will be ~0.1 arcsec within three-sigma. The bus provides 300 W of electric power for all instruments. The data downlink rate can reach 4 Mbit/s. The WSO-UV will be launched from Baikonur (Kazakhstan) with a Proton rocket [16]. The geosynchronous orbit offers several advantages: stability of the orbit, continuous visibility zones, minimal duration of Earth shadow periods, and a continuous link between space and ground segments for radio communication. Earth occultation periods will be short enough so that the WSO-UV orbital period will allow long term monitoring and rapid access to targets of opportunity. The radiation belt, however, can affect the detector backgrounds. Thanks to the inclination of 40°, the radiation flux at high latitude regions is low enough, about five orders lower than that at the equatorial region; however observation periods must still be planned according to the radiation background.

The two principal science instruments are located in two compartments: a field camera unit (FCU) and WSO-UV spectrographs (WUVS), as shown in Fig. 1. The FCU contains two imaging cameras: the FUV camera is sensitive to far-UV starting from ~115 nm and the NUV camera is sensitive to near-UV and to the visible wavelengths. The WUVS contains three spectrographs: a UV Echelle spectrograph (UVES), a vacuum UV Echelle spectrograph (VUVES), and a long-slit spectrograph (LSS). Detailed information can be found in [17]. The geometrical stability of the spectrograph is provided by titanium heat pipes instead of the previously planned CeSiC material [18]. All heat sources inside spectrographs are connected to the outer wall by additional heat pipes. The cooling system for the detectors consists of a large external radiator and isolated heat pipe connected to detector cold fingers [19]. Stellar spectroscopic studies that will involve WSO-UV are discussed in [20, 21].

The most advanced unit of the mission is the telescope, which has already passed its vibration tests successfully, as well as a number of key subsystems, such as the guidance system, science computer, etc.

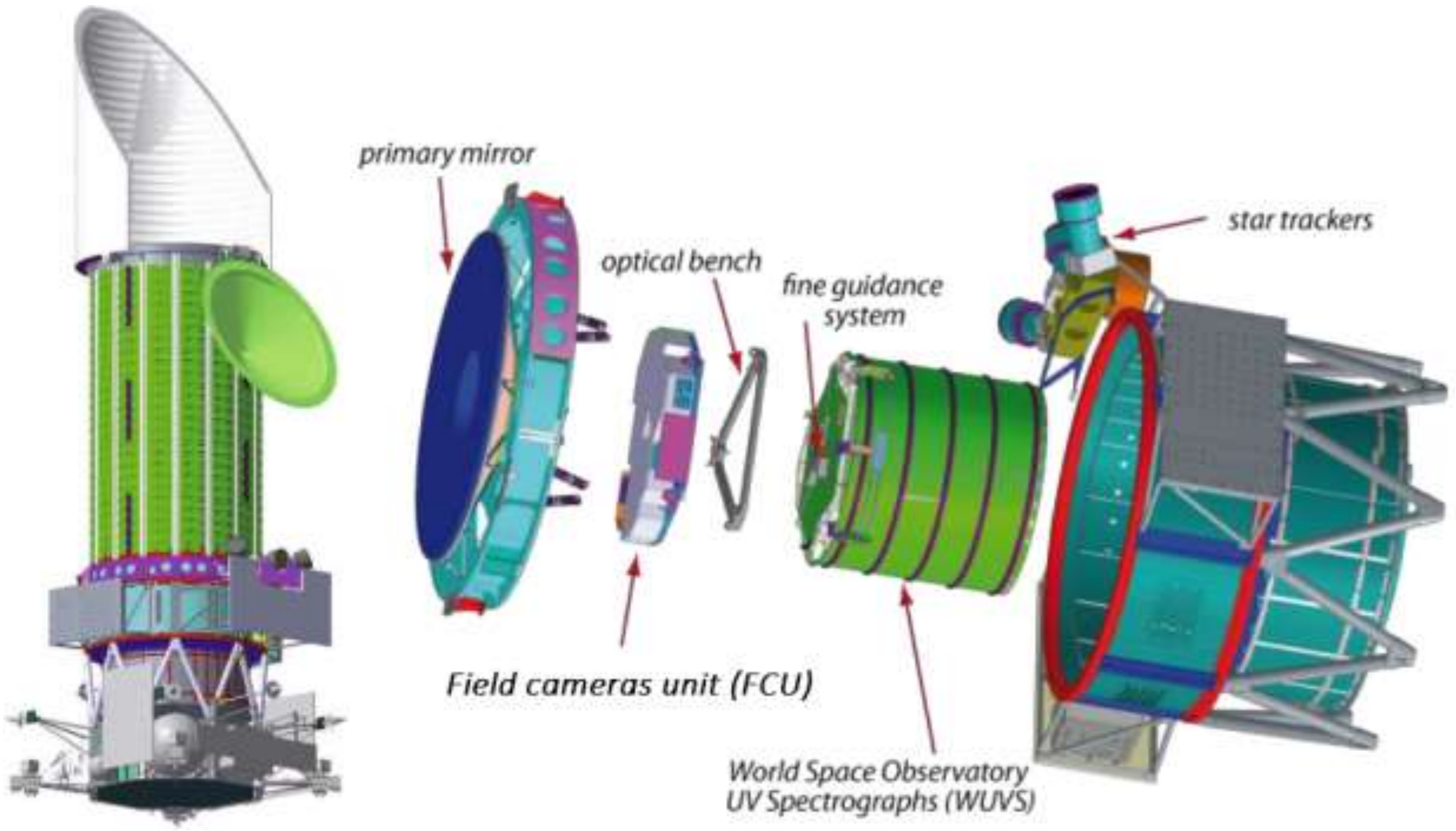

**Fig. 1** The WSO-UV orbital telescope design view (left) with instrumental compartments shown (right).

## 3 Science Goals

### *3.1 WSO-UV SCEDI Coronagraph: Aims, Resolution, and Contrast*

The SCEDI coronagraph will attempt to directly image more massive and more distant planets from their host stars. The direct detection of Earth-like planets in the habitable zone is not the ambition of the SCEDI coronagraph.

The capabilities of several stellar coronagraphs aimed to study exoplanets are shown in Fig. 2. Compared to the ground-based Subaru telescope (8.2 m primary mirror) with HICIAO coronagraphic instrument (used in the SEEDS program [ 22 ]), and to the state-of-the-art WFIRST-AFTA coronagraphic mission planning to use the 2.4 primary meter of surplus telescope from the National Reconnaissance Office (NRO) with CGI instrument [7], the SCEDI instrument onboard the WSO-UV 1.7 m telescope proposed in this paper can achieve a similar resolution as an 8 m ground-based telescope in an observing mode dedicated for dim exoplanets. Moreover space telescopes have the potential to achieve deeper coronagraphic contrast.

A science goal for the stellar coronagraph onboard the WSO telescope is to directly image exoplanets and circumstellar discs by searching 20–30 stars near our Solar system. Planets and discs can be observed at multiple epochs down to 7–9-th order peak-to-peak (star-to-planet) coronagraphic contrast, which enables detecting and characterizing exoplanets down to super-Earth sizes. As a minimum success scenario, the goal is to test the coronagraph as an in-flight demonstrator, and to photometrically characterize ~10 giant planets, which have been previously detected by the radial velocity technique.

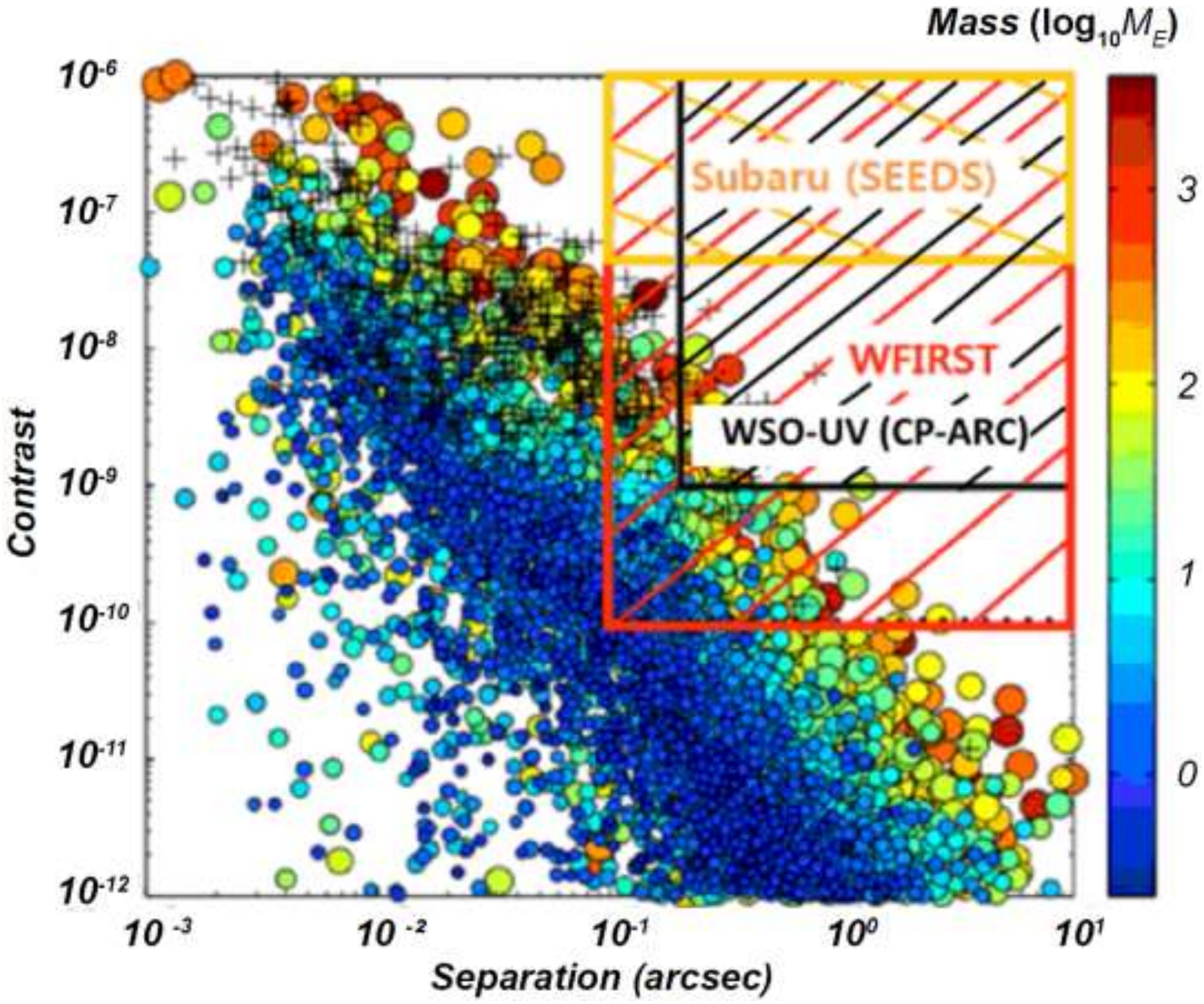

**Fig. 2** The known exoplanets distribution, the vertical axis is the logarithmic contrast (the ratio of luminosity of the host-star to exoplanet), the horizontal axis is the apparent separation between the planet and the host star, and color shows the exoplanets mass in logarithms of M⊕. Rectangles show the observing capabilities using the HICAO coronagraph of the 8.2 m Subaru ground-based telescope (orange rectangle), the CGI coronagraph onboard the future WFIRST-AFTA 2.4-m space-based telescope (red rectangle), and via the proposed SCEDI coronagraph onboard the WSO-UV 1.7 m telescope (black rectangle).

A CP-ARC – common-path achromatic rotational shear coronagraph [11, 12], is planned (for the first time) as a coronagraphic device for real onboard observations, and it has a science-related advantage. CP-ARC introduces a centrostellar occultation effect under a variable angle of rotational shear. A rotational shear interferometer (RSI) superposes a telescope pupil and its rotated copy with an achromatic antiphase and results in destructive interference. If the angle of the rotational shear is rather small (fixed between 5 and 10 degrees), the nulling interferometer demonstrates reduced sensitivity to low-order aberrations from telescope optics. Because of WSO-UV requirements, a stellar coronagraph should be a simple instrument without the implementation of any adaptive optics system to correct the wavefronts. CP-ARC will be able to image an exoplanet by balancing the raw coronagraphic contrast and the spatial resolution (inner-working angle – IWA). Considering such an instrumental capability, we studied the CP-ARC in detail (Sect. 4).

*3.2 UVSPEX Spectrograph Science Goals*

PROCYON with a Lyman Alpha Imaging CAmera Cassegrain telescope has observed geocorona images [13]. Images are formed from exospheric atoms observed from up to more than 240,000 km, which is about 38 Earth radii. The hydrogen density is estimated at 20 atoms/cm$^3$ at a distance of ~60,000 km in the Earth's exosphere. The same density is expected to be observable at distances of 10,000–20,000 km in Venus and 30,000–35,000 km in Mars. For Mars, the exosphere height can be even lower, and is caused by the difference of the mixing ratio of $CO_2$ in the upper atmosphere. Venus and Mars have $CO_2$-rich atmospheres with lower exospheric temperatures [23]. On Earth, $CO_2$ is removed from its atmosphere by a carbon cycle involving the oceans and tectonics [24]. Translating these arguments to exoplanets within the habitable zone presents a possible marker to distinguish an Earth-like planet from a Mar-like or Venus-like planet. The expanded exospheres can be observed in UV during the exoplanet transit event in a primary eclipse, which reduces the stellar flux when an exoplanet passes in front of the host star.

Moreover, theoretical exospheric models [23, 25] extrapolate these arguments to an oxygen exosphere. There are three oxygen emission lines (O I-triplet), the wavelengths of which are 130.2 nm, 130.5 nm, and 130.6 nm. Fig. 3(a) shows the observed spectrum of Proxima Centauri with spectral range from 130 nm to 131 nm, and Fig. 3(b) shows it from 130.57 nm to 130.62 nm [26]. Although we can see Earth's oxygen emission lines in these results, they would be negligible using WSO-UV because of its high altitude.

Note that all other emission lines, e.g., emission line around 130.3 nm, were taken into account in the following calculation, while we excluded the 130.2 nm O I line (at which the emission is caused by resonance scattering from oxygen at the ground state), to avoid uncertainty in the cold interstellar oxygen absorption.

Then for the oxygen line, the predicted transit depths are shown in Fig. 3(c) evaluating whether the Proxima Centauri b exoplanet [27], at which the EUV irradiation is much higher than Earth's, is either Earth-like, Mars-like, or Venus-like. The temperature of the upper atmosphere for Earth-like exoplanets is estimated to be ~10,000 K [25]. In contrast, those for Mars-like and Venus-like exoplanets are estimated to be ~300 K and ~600 K, respectively [23]. The transit depth at the line center for each case is 76%, 0.7%, or 3.8%, respectively. Owing to the large difference in the transit depth, Earth-like planets can potentially be distinguished from Venus-like and Mars-like planets [28]. High dispersion would be required to resolve the absorption feature in the O I line shown in Fig. 3(c). However, the total photometric transit depth of stellar emission integrated from 130.25 nm to 130.75 nm corresponds to 25%, 0.11%, or 0.20%, respectively, which are also distinguishable. As shown in Sect. 6, we selected a low-dispersion, high-efficiency design to observe M stars dark in the UV region and to simultaneously detect both hydrogen (~122 nm) and oxygen (~130 nm) emission lines. Note that hydrogen can be observed only when the relative (heliocentric) velocity of the exoplanetary system is high enough to avoid critical interstellar absorption. For example, the interstellar absorption of the hydrogen emission line at 12 pc is ~100% if the relative velocity is < 30 km/s; however, the transmittance is >~25% if the relative velocity is >50 km/s [29]. According to the RECONS (Research Consortium on Nearby Stars) catalog ( [30]), at least in the range of <5 pc, there are seven M dwarfs with relative velocities > 50 km/s, which will potentially be our observation objects.

In addition to exospheres, stellar UV, especially H-Lyman alpha emission, has a significant effect on photochemistry in the lower atmospheres of exoplanets. Recently, multi-wavelength transit observations have been performed in the IR and optical bands for constraining the properties of exoplanet atmospheres from wavelength-dependent transit depths. This type of observation is often called transmission spectroscopy. Although the number of examples is small, it has been already revealed that transmission spectra of close-in exoplanets are diverse [31]. In particular, a likely candidate for the diversity in transmission spectra of close-in exoplanets is photochemically produced clouds, which are often called haze [32]. Indeed, transiting planets that have atmospheres with hydrocarbon haze are theoretically demonstrated to show diverse transmission spectra, depending on UV irradiation levels [33]. As the UV emission of low-temperature stars (i.e., M dwarfs) varies considerably with time, simultaneous observation in the UV and IR/optical range is crucially important and would yield valuable clues to the origin of the diversity, which is quite uncertain at present.

To summarize the UVSPEX science goals:

1. To distinguish an Earth-like planet from Venus-like or Mars-like by observation of its oxygen and hydrogen exosphere.

2. UV irradiation intensity, especially H-Lyman alpha flux, is important for estimation of the extension of the exosphere and also for photochemistry analysis in the lower atmospheres.

To realize exoplanet transit observations in oxygen spectral lines with the desired accuracy, we will equip the WSO-UV telescope with the UVSPEX spectrograph.

Science and technical requirements are discussed in Sect. 5.

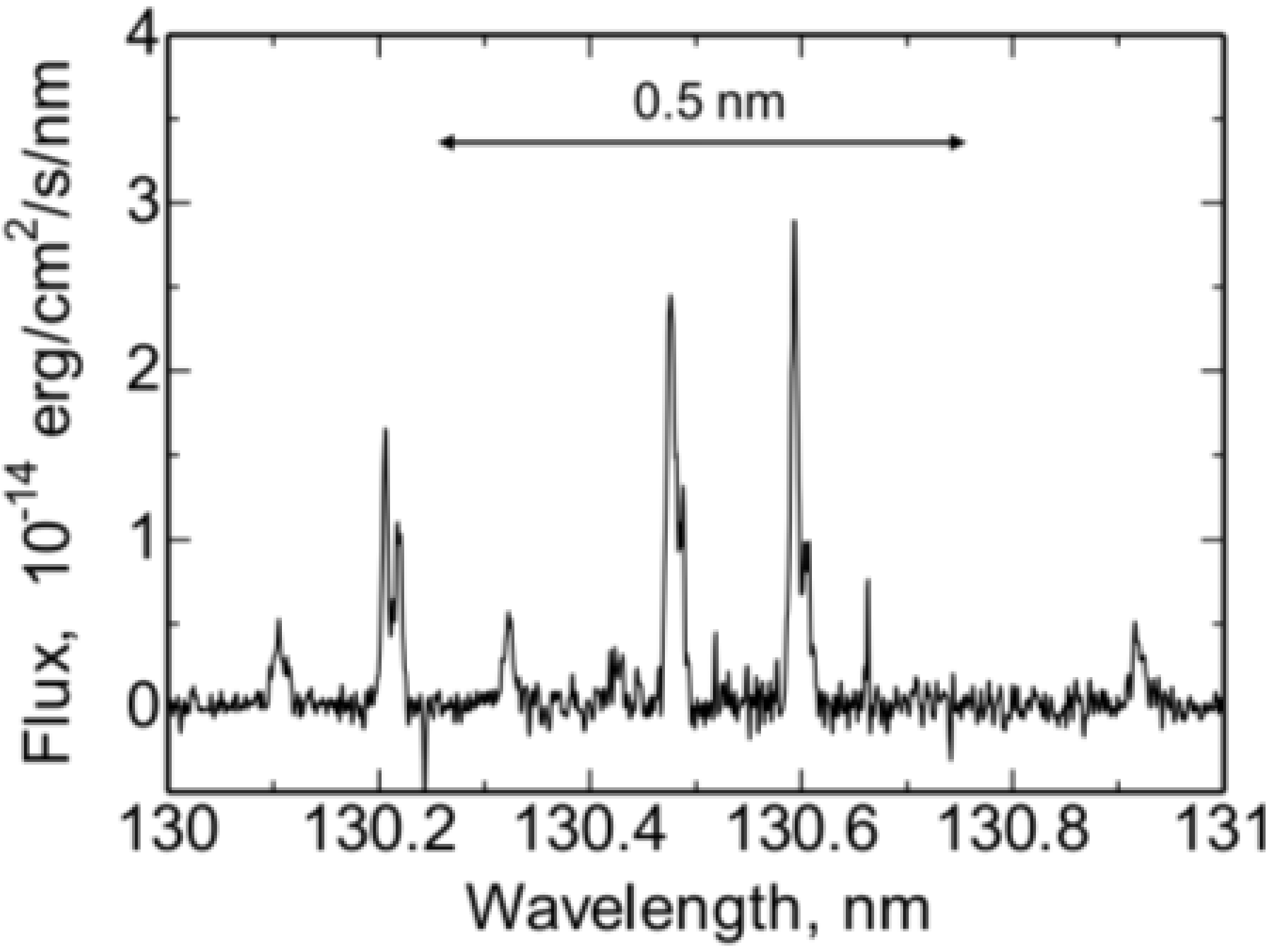

(a)

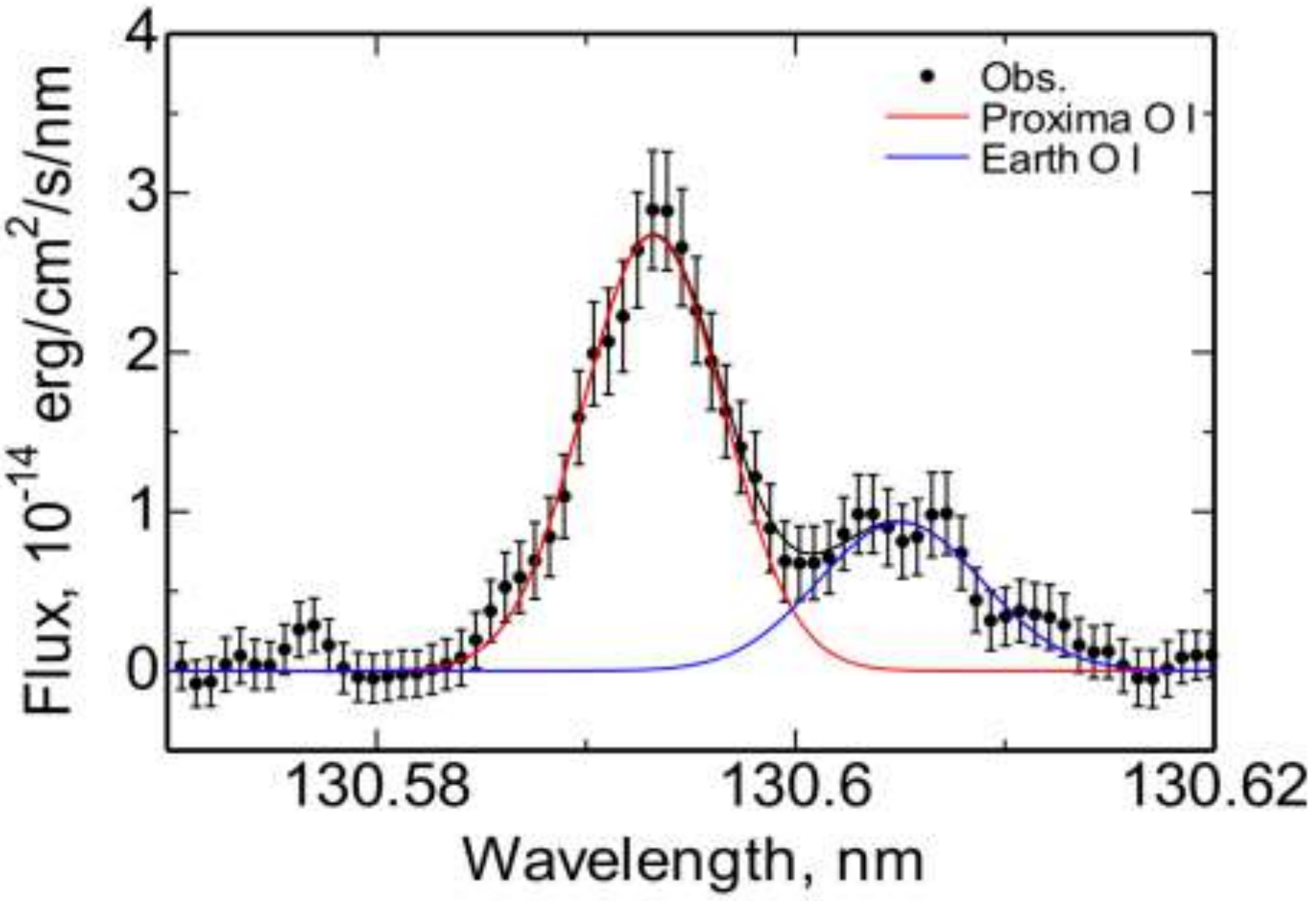

(b)

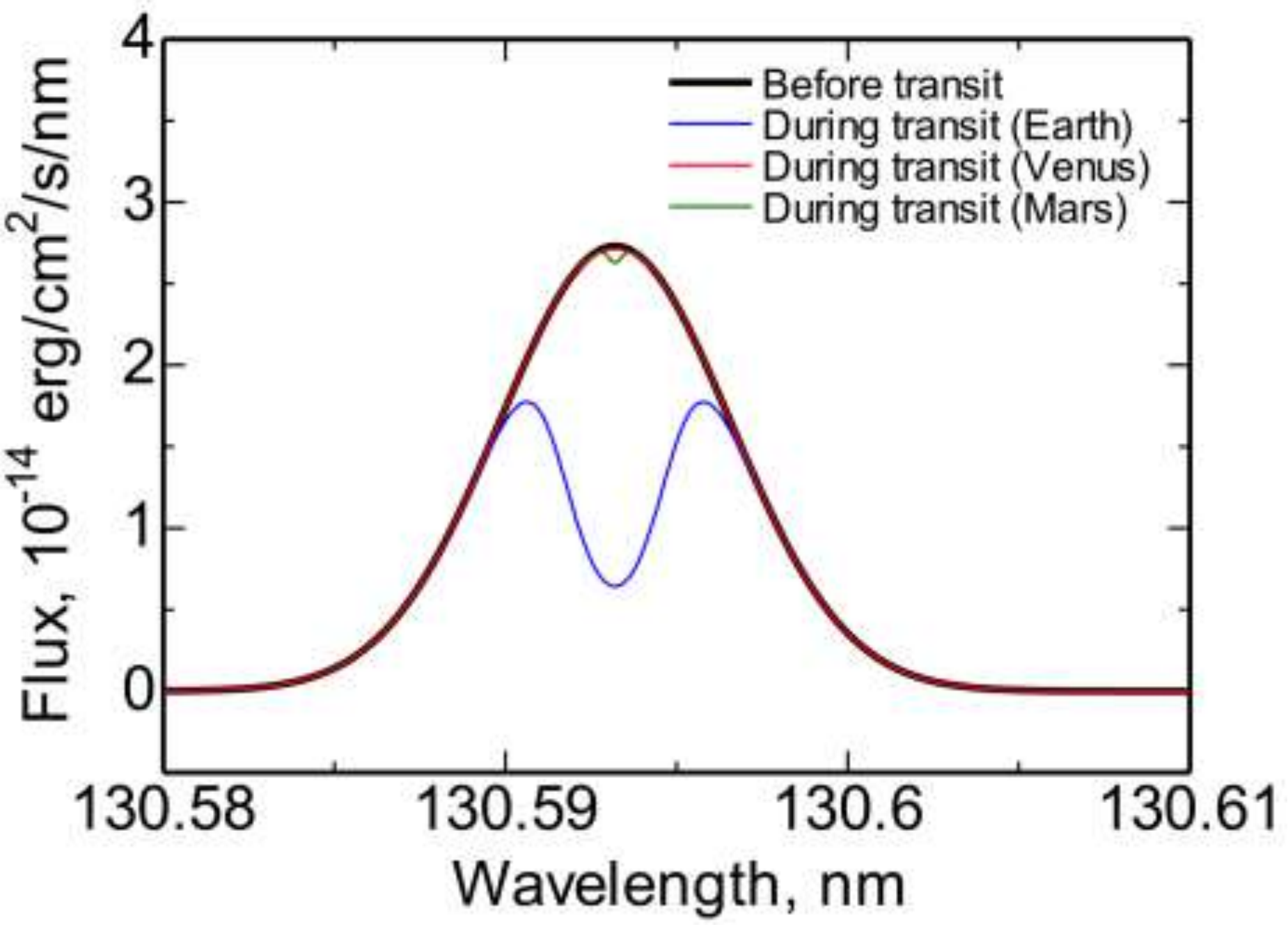

(c)

**Fig. 3** (a) – The spectrum of Proxima Centauri obtained by HST/STIS. The arrow shows the spectral range in which stellar emission intensity was integrated. (b) –Spectrum in the spectral range around O I 130.6 nm. The red curve shows the result of Gaussian fitting for the emission line of Proxima and the blue shows that of the Earth's oxygen corona. (c) – Predicted theoretical transit photometric curves of the Proxima Centauri b exoplanet in oxygen OI spectral line at 130.6 nm for Earth-like (blue), Venus-like (red), and Mars-like (green) in comparison with the non-transit curve (black).

## 4 Coronagraph Instrument SCEDI: Technical Aspects, Requirements and Implementation

### *4.1 Wavefront Quality Requirements for a Stellar Coronagraph*

A stellar coronagraph instrument on a space telescope requires wavefront correction with extremely high precision. For the imaging of an Earth-like exoplanet, a wavefront with $\lambda/10^4$ rms precision (and amplitude inhomogeneity of $1/10^3$) is required to achieve a desired $10^{9\text{-}10}$ coronagraphic contrast, for proximity to the telescope resolution maximum (of 1-2 $\lambda$/D) [34]. A stellar coronagraph with an external occulter, such as *TPF-O*, and more ambitious projects to image exoplanets [35], do not require high wavefront quality (close to ideal), but they do require spacecraft in difficult orbits.

Present day technology allows manufacture of the telescope primary mirror with a surface figure accuracy not exceeding $\lambda$/10 p-v. This measure characterizes aberrations with low spatial orders. At mid- and high- spatial frequencies, the measure of power spectral density (PSD), which characterizes the level of light scattering, can show much better quality than the mean surface roughness (rms).

By means of adaptive optics (AO) by deformable mirror and by wavefront sensor, one can provide a closed-loop wavefront control within the desired precision ($\lambda/10^3 \ldots \lambda/10^4$). In a remote space experiment, such a wavefront control unit requires additional volume, load, and onboard computational facilities. Nevertheless, the WFIRST-AFTA CGI is going to implement an extremely precise wavefront correction (ExAO) onboard.

In our proposal for the SCEDI coronagraph onboard the WSO-UV orbital telescope, we shall not use ExAO owing to simplification requirements and programmatic considerations. However, mid- and high- spatial frequencies of the telescope optics have small surface errors, so WSO-UV's telescope can work in the visible spectral band, where optical surface errors and scattering are reduced. Because of the requirement to have a small amount of scattered light in the wavelengths of vacuum UV (from 104 nm), the PSD in the mid- to high- spatial frequencies at visible wavelength $\lambda$=632.8 nm is prescribed below 1 nm rms.

In further analysis, we aim to propose an engineering solution for the coronagraph instrument under the conditions of existing telescope optics: a cumulative figure error of $\lambda/5 - \lambda/10$ p-v in the low spatial frequency domain and microroughness of 1 nm rms in a spatial frequency domain of 1 cm (100 cycles/m).

### *4.2 Principle of Achromatic Interfero-Coronagraph with Small Rotational Shear*

Among the variety of stellar coronagraphs [ 36 ], the working principle of an interfero-coronagraph is to superimpose (with antiphase) two pupil images being shifted, being swapped, or being rotated. Additionally, this results in destructive interference and therefore eliminates starlight received from the on-axis. Planetary (or companion) light does not interfere destructively because the off-axis tilt (or shift in pupil) separates the two companion copies spatially.

Interfero-coronagraphy is advantageous in terms of its broad achromaticity and, more generally, because of its small inner working angle (IWA). The AIC coronagraph is known to have one of the smallest possible IWAs, and it was first proposed in [37] with a fixed angle of rotational shear of 180°. Later, it was re-designed using a modified Sagnac scheme, implementing the common-path (or cyclic path) (CP-AIC) [38] to relax the original mechanic instability.

A disadvantage of interfero-coronagraphy, known as stellar leakage effect, is that a star's light is not fully suppressed because of the apparent size of the star [36]. The observed star size is not directly resolvable; however, physically it is not a spatially coherent point source. Observed by means of a two-meter diameter telescope (e.g. by the WSO-UV), the Solar-type star at 10 parsec has an apparent size of ~0.02·λ/D, limiting the destructive interference to 4–5 orders of magnitude for starlight suppression. To overcome this effect, an interfero-coronagraph with variable rotational shear, abbreviated as ARC [39] (later re-designed in a mechanically robust common-path (or cyclic path) CP-ARC [11, 12]), offers the possibility to adjust the arbitrary pupil rotation angle to an optimal angle for the observing configuration. With a 1.7 m WSO-UV telescope, the 10 degree image rotating CP-ARC suppresses stellar leakage caused by a finite stellar disk for a Sun-like star at 10 pc, almost eight orders of magnitude below the raw coronagraphic contrast. This is a theoretical contrast without any aberrations.

In principle, a smaller rotational shear angle (5–10°) is characterized by a worse IWA and therefore a worse spatial resolution, as compared to the 180° case. With reduced rotational shear, the IWA increases and the spatial resolution decreases. Therefore, the coronagraph will have reduced capability to detect the companion in close proximity to the star. Fig. 4 illustrates this dependency for the rotational shear angles of 180°, 10°, and 5°.

The throughput, $T(\rho_0)$ shown in the graphs was evaluated from

$$T(\rho_0) = 1 - \frac{2J_1\left(\frac{2\pi}{\lambda} D\rho_0 \sin\left(\xi/2\right)\right)}{\frac{2\pi}{\lambda} D\rho_0 \sin\left(\xi/2\right)} \tag{1}$$

(Eq. (13) in [39]), and describes the throughput of an interfero-coronagraph for a companion signal versus the observed centrostellar position $\rho_0$ (in λ/D units) of the companion for different rotational shear angles ξ.

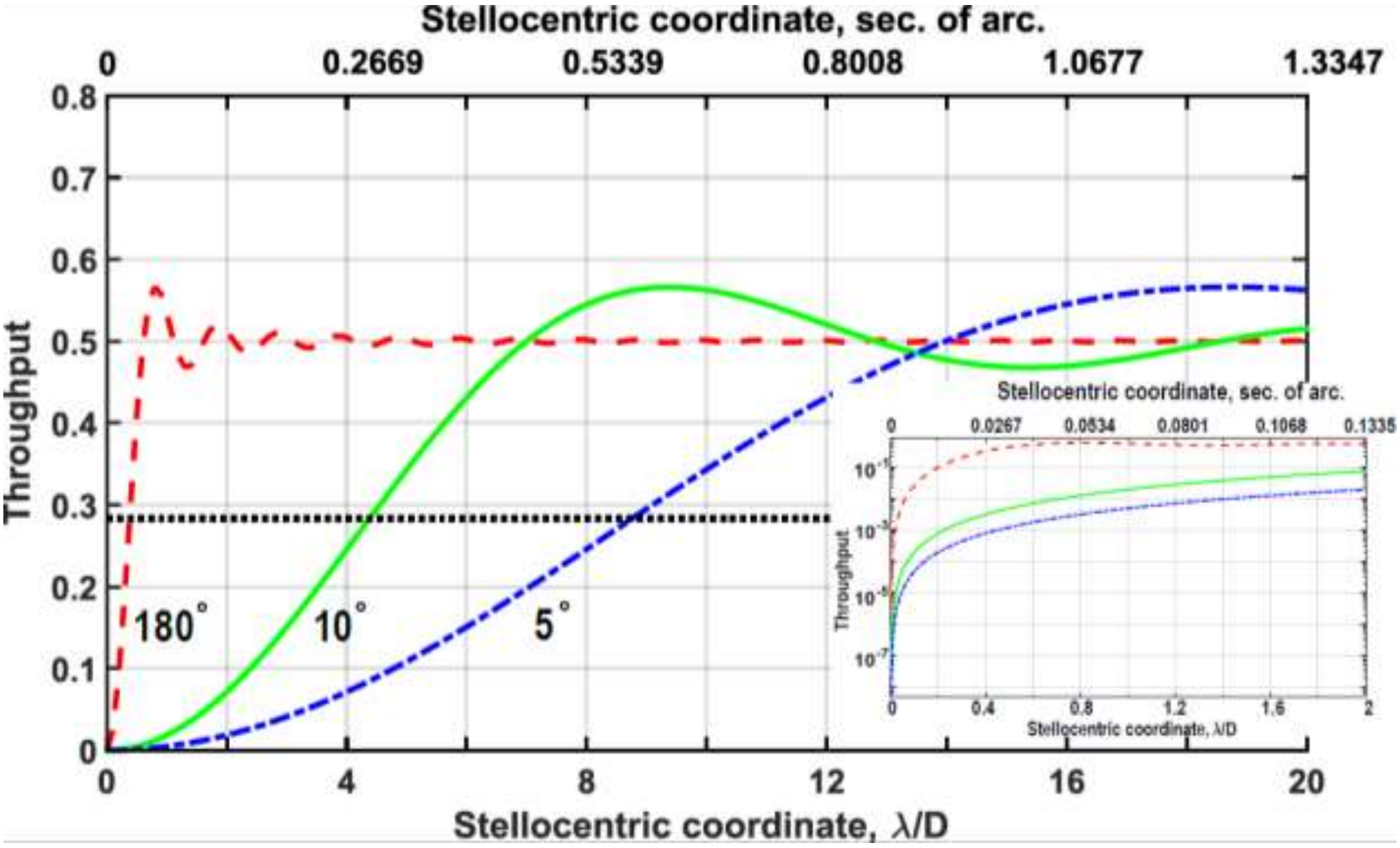

**Fig. 4** Theoretical throughput of an interfero-coronagraph versus the apparent stellocentric separation of a companion (Eq. 1) evaluated for three different rotational shear angles. Upper (angular) axis corresponds to the WSO-UV telescope operating at the central wavelength of 550 nm. 180º – dot-dashed blue curve, 10º – solid green curve, 5º – dashed red curve. The coronagraph IWAs at the considered angular shears are given as half-maxima (horizontal dot-marked black line) of the throughput curves: 180° – 0.38·λ/D, 10° ~ 4.5·λ/D, and 5° ~ 9·λ/D.

However, reducing the rotational shear angle to smaller values makes CP-ARC much less sensitive to the residual wavefront errors coming from the figure errors of the primary and secondary telescope mirrors. Practically, the interferometer visualizes the incident wavefront with the lower sensitivity. This feature of rotational shear interferometers (RSI) is described in [40, 41]. As a result, an interfero-coronagraph with angular shears of 5–10° can achieve a significant coronagraphic effect without extreme wavefront correction. This effect can be realized within the specifications of the WSO-UV telescope, with the primary and the secondary mirrors together having optical aberrations of ≤λ/5 at λ =633 nm (at low spatial frequencies) and with the microroughness at the mid spatial frequency of ≤1 nm rms (at a 1 cm scale).

In Fig. 5, we show several simulation results for a classic interfero-coronagraph scheme to demonstrate (for the WSO-UV specifications) an almost disabled coronagraphic effect with a 180° rotational shear, and the enabled coronagraphic effect with 5° and 10° rotational shears. To simulate microroughness, the power spectral density (PSD) has been modeled from actual measurements that follow a slope approximately proportional to $\sim e^{-11/3}$. For the low order spatial frequencies aberrations, we tested several random Zernike coefficients up to the Zernike polynomial order of 36, with the constraint not to exceed λ /5 p-v at 633 nm. Variability of about a half order in raw contrast units was evaluated. Microroughness was considered by PSD in the range of mid- and high- spatial frequency aberrations. Microroughness estimated by rms was set at 1 nm within the characteristic (cut-off) length of 1 cm (100 cycles/m). Variation of the exponential slope was studied from the declared -11/3≈-3.7 down to -2.17 (known for the HST [42]).

The raw coronagraphic contrast can reach a $10^6$–$10^7$ peak-to-peak value at 7–10 λ/D stellocentric distances. In Fig. 5, the coronagraphic contrast was estimated by comparing a non-coronagraphic point spread function (PSF) and a coronagraphic PSF. The PSF dependencies were azimuthally averaged.

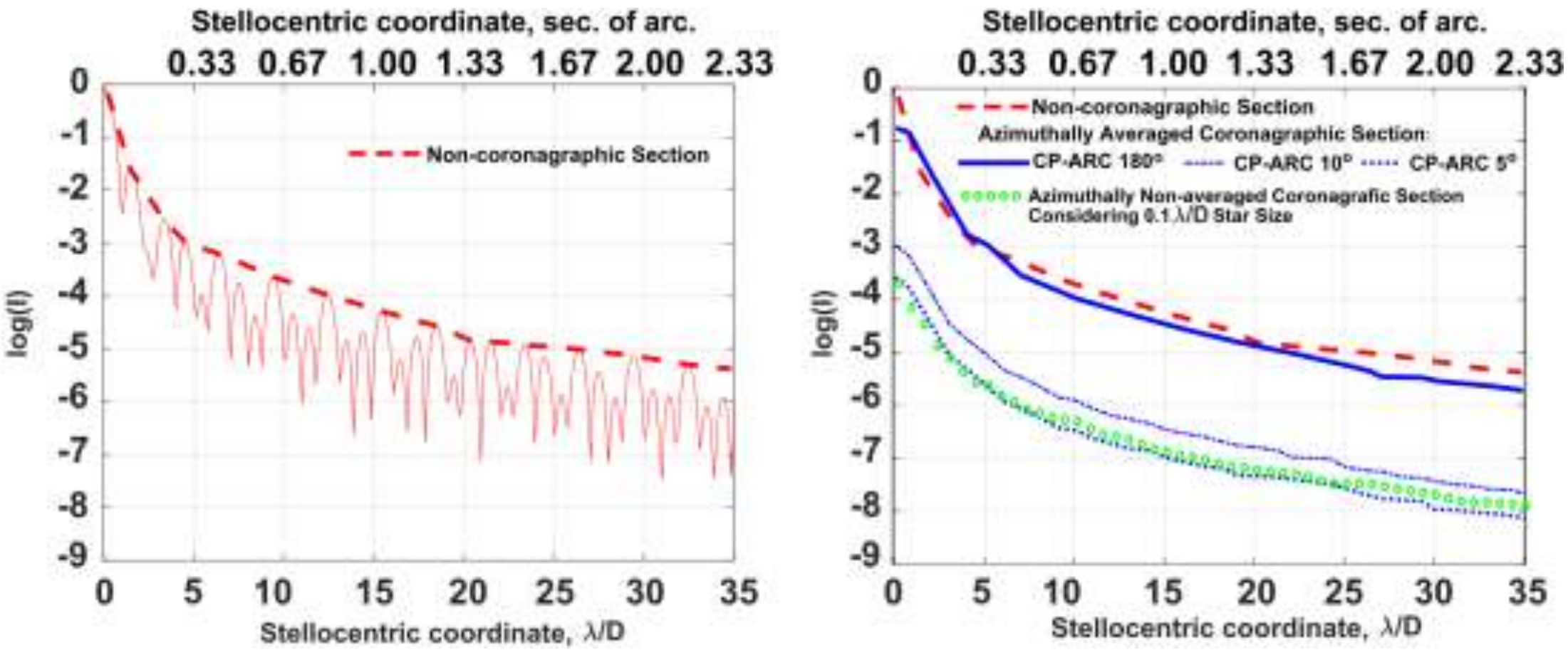

(a)

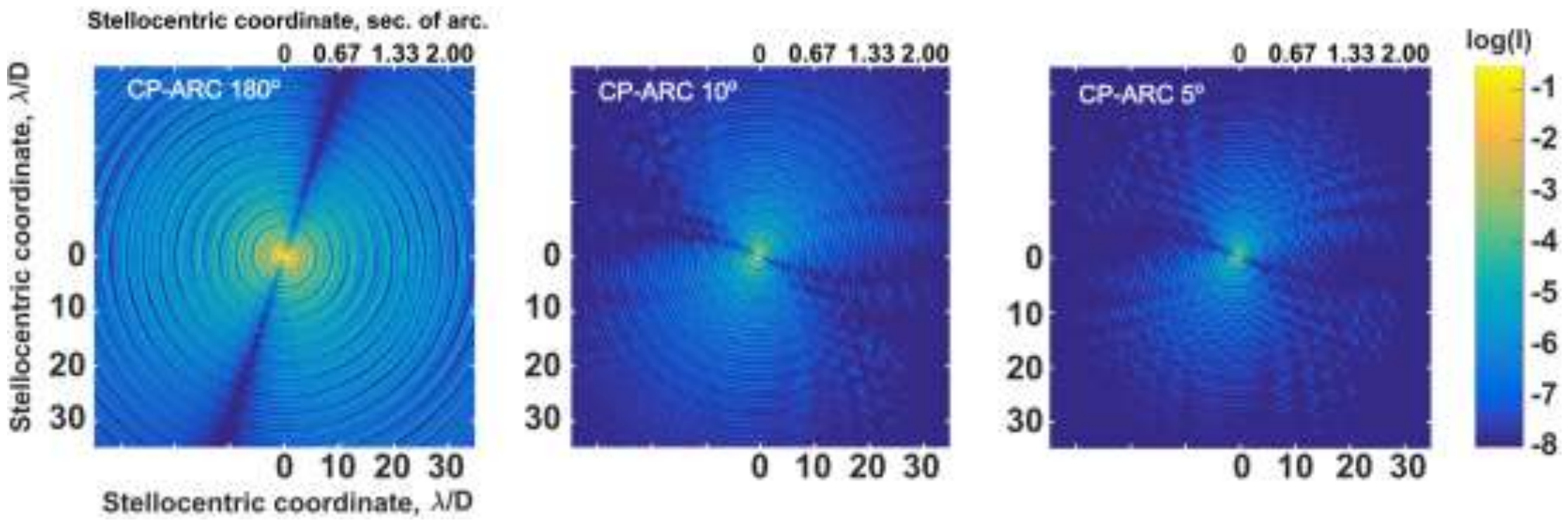

(b)

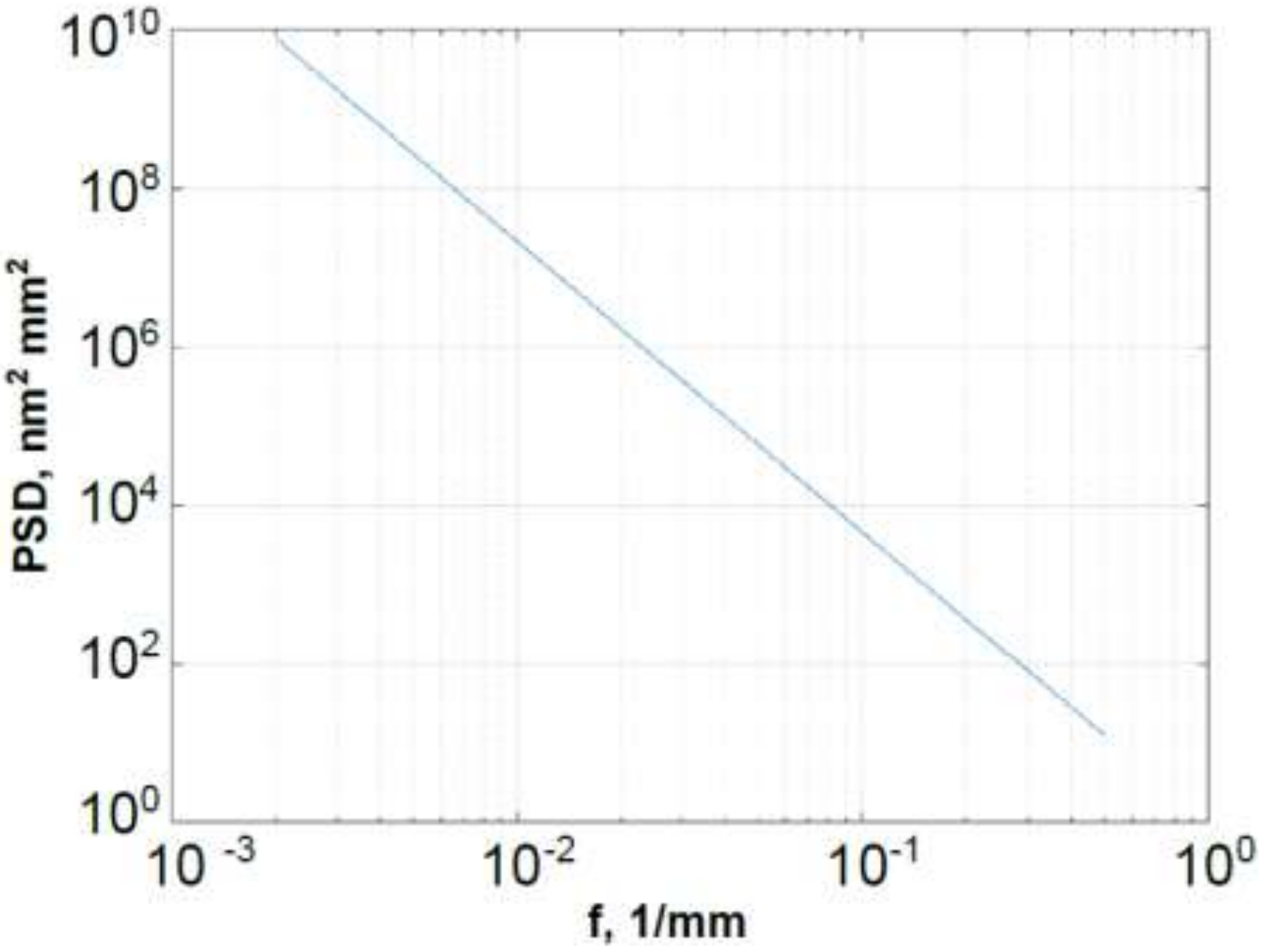

(c)

**Fig. 5** Coronagraphic (raw) contrast achievable by means of interfero-coronagraph with the rotational shears. (a) – Non-coronagraphic PSF (left). PSF envelopes (right): non-coronagraphic (red dashed) and coronagraphic 180º (blue solid), 10º (blue dashed), 5º (blue dotted), evaluated for the WSO-UV telescope specs: λ/5 p-v low-order aberrations and 1 nm rms at mid frequencies of PSD ~ $e^{-11/3}$, at λ=550 nm. Green o-marked line stands for 0.1 λ/D stellar size and is affected by the considered aberrations and microroughness. (b) – 2D coronagraphic raw contrast maps for 180º, 10º, and 5º rotational shears (from left to right). (c) – PSD approximated from primary mirror test data.

We have estimated the effect of an extended source considering a 0.1 λ/D stellar size, aberrations, and micro-roughness. Without the two latter factors, theoretically [12, 39], the coronagraphic contrast is predicted to be better than 7 orders at a stellocentric distance 10 λ/D. Here the effect of scattering compared to optical low-order aberrations and surface micro-roughness becomes dominant. In Fig. 5 (a) on the right panel we added the corresponding graph (green line, o-marked) for 0.1 λ/D stellar size affected by λ/5 p-v low-order aberrations and 1 nm rms represented by the PSD ~ $e^{-11/3}$.

*4.3 Coronagraph Sensitivity to the Pointing Error*

The proposed interfero-coronagraph CP-ARC with 10º rotational shear is also less sensitive to the pointing error, as compared to the interfero-coronagraph having 180º rotational shear. The requirement on the pointing error of the WSO-UV fine guidance system is 0.1 arcsec ($\approx 1.5\cdot\lambda/D$) within $3\sigma$. We examine here how critical is this nominal 0.1 arcsec pointing error for the operations of the coronagraph instrument.

In Fig. 6, we can study the degradation of the raw coronagraphic contrast at stellocentric separation, where a companion is expected to be observed, versus the pointing error. For the pointing error of 0.1 arcsec, the raw contrast at $10\cdot\lambda/D$ ($\approx$0.67 sec. of arc) is shown as nearly twice degraded.

The factor of two is not a critical issue. However, the central part of the coronagraphic image detector, which is optically aligned on a centrostellar axis, receives an increased photon flux from stellar leakage because of the pointing error. So, the dynamical range of the detector is reduced and it may become saturated, which is a critical issue to be corrected.

The detector for SCEDI is composed of MCPs and a resistive anode, which is identical to that in UVSPEX (see Sect. 5) except for the photocathode material. The detector plane of SCEDI has 512x512 pixels, includes an InGaAs photocathode for the visible spectral range. The maximum count rate is ~20,000 cts/s for whole region of detector and the dark noise is <<0.01 cts/s/pix. The photon flux from the obscured star was estimated, it allows us to use the detector dynamical range to several orders for post-processing.

To preserve the dynamical range of the photon counting detector that can be dramatically affected by the strong stellar leakage caused by the pointing error, we decided to place an opaque mask in a relay focal plane, just before the detector. This mask was chosen to be ~6 $\lambda/D$ in diameter as a compromise between the coronagraph sensitivity in that range and the amount of leakage.

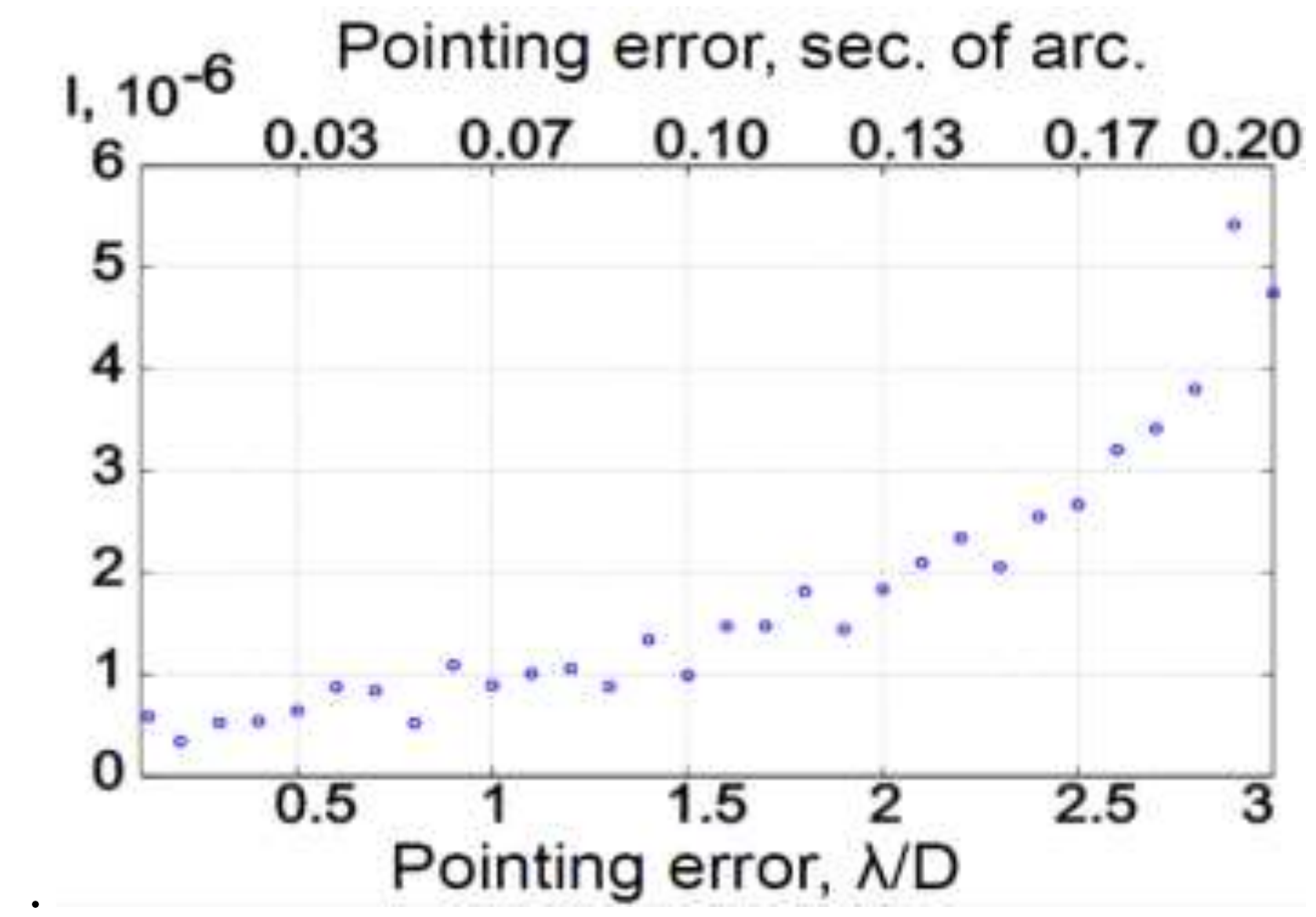

**Fig. 6** Coronagraph sensitivity to pointing error – the (raw) coronagraphic contrast at 10·λ/D stellocentric position versus the pointing error.

### *4.4 Influence of the Telescope Spider Mask*

In the proposed coronagraph SCEDI 10º CP-ARC, one more source of stellar leakage to be evaluated is caused by diffraction from the telescope spider arms. The WSO-UV telescope has a four-arm symmetric spider holding the Ø 370 mm secondary mirror. Each arm is 24 mm wide. With unblocked diffracted light from the spider arms, the coronagraphic image shows a raw contrast of ~$10^5$ at 10·λ/D, nearly two orders lower than it would be without the spider.

In Fig. 7(c), we plot the non-coronagraphic PSF (red line) and the coronagraphic PSF (blue line). We note that an insufficient level of the non-coronagraphic PSF signal reduction at the desired 7–10 λ/D stellocentric positions is achieved. To verify the origin of the leaked light caused by the spider, we can study the plots in Fig. 7(a) of two coronagraphic images, one obtained in the secondary pupil (shown on the left) and another (focused) in the image plane (shown on the right), where most of the light definitely originates from the spider arms. The CP-ARC interfero-coronagraph superposes two pupil images with a rotational shear. The spider arms are imaged as doubled by the angular shear, and they are mostly bright.

To cancel the diffraction from the spider arms, we re-image the spider plane by a focusing mirror (e.g., F'~200 mm calculations) in the plane optically conjugated with the spiders' front plane (different to the secondary pupil). This assigns a plane, different to a classical Lyot-stop plane, conjugated to the pupil. Then, we simulated an opaque mask of the form shown in Fig. 7(b). This mask hides the image of spider arms using 2.0 mm-wide stripes. We then re-focus the resultant image in the photo-detector plane, where we study the filtered coronagraphic PSF.

The coronagraphic image with the proposed spider mask demonstrates the enhanced raw coronagraphic contrast (down to between $10^{-6}$ and $10^{-7}$), which is similar to the case without the spider arm effects, and is comparable to Fig. 5.

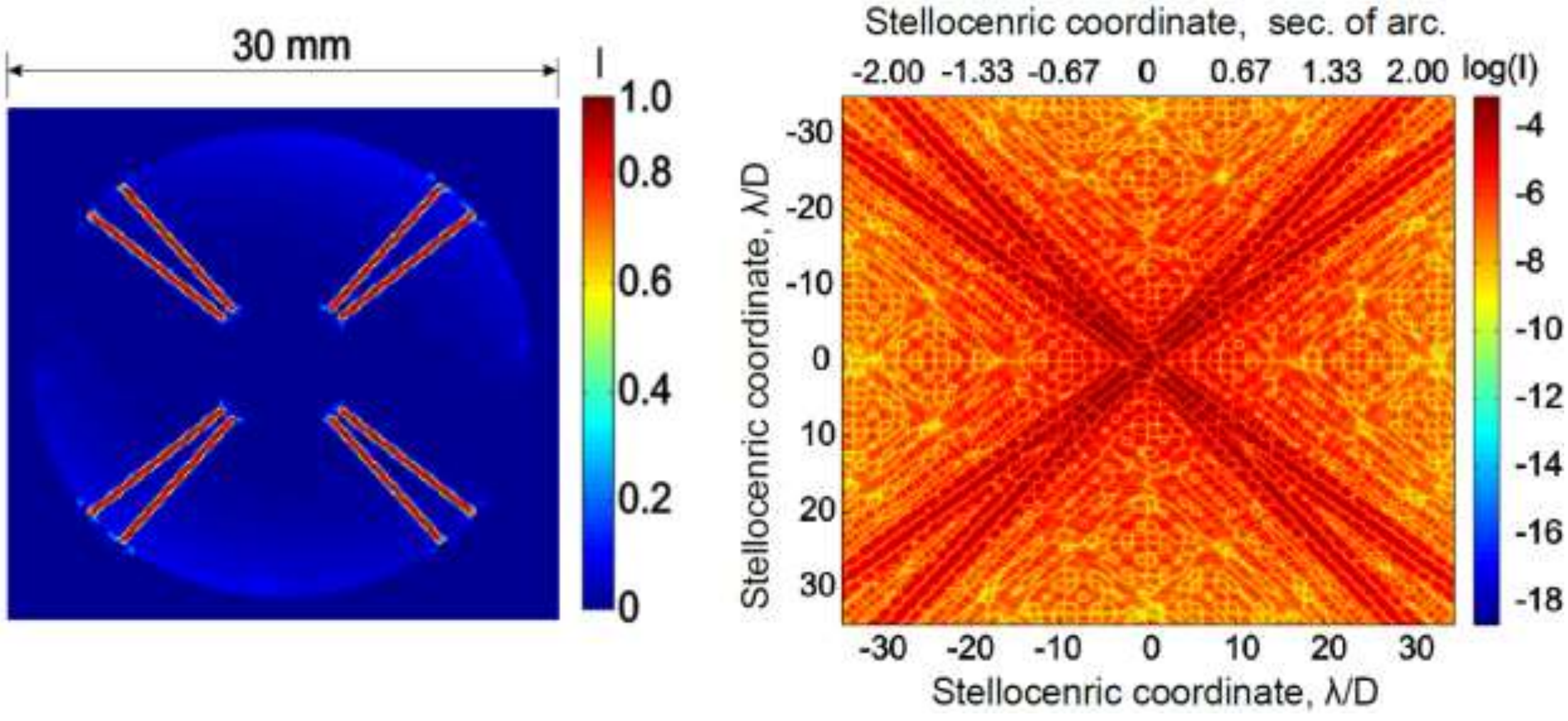

(a)

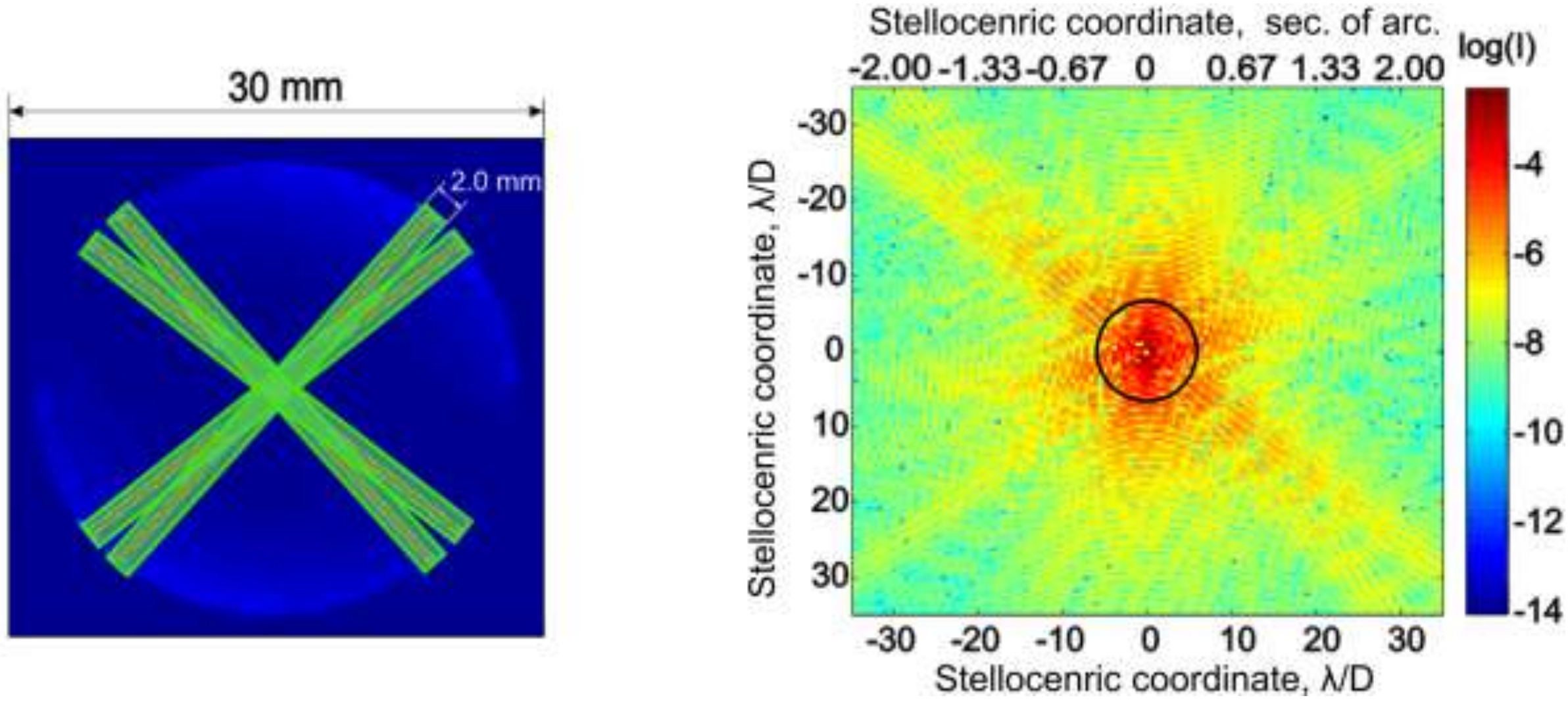

(b)

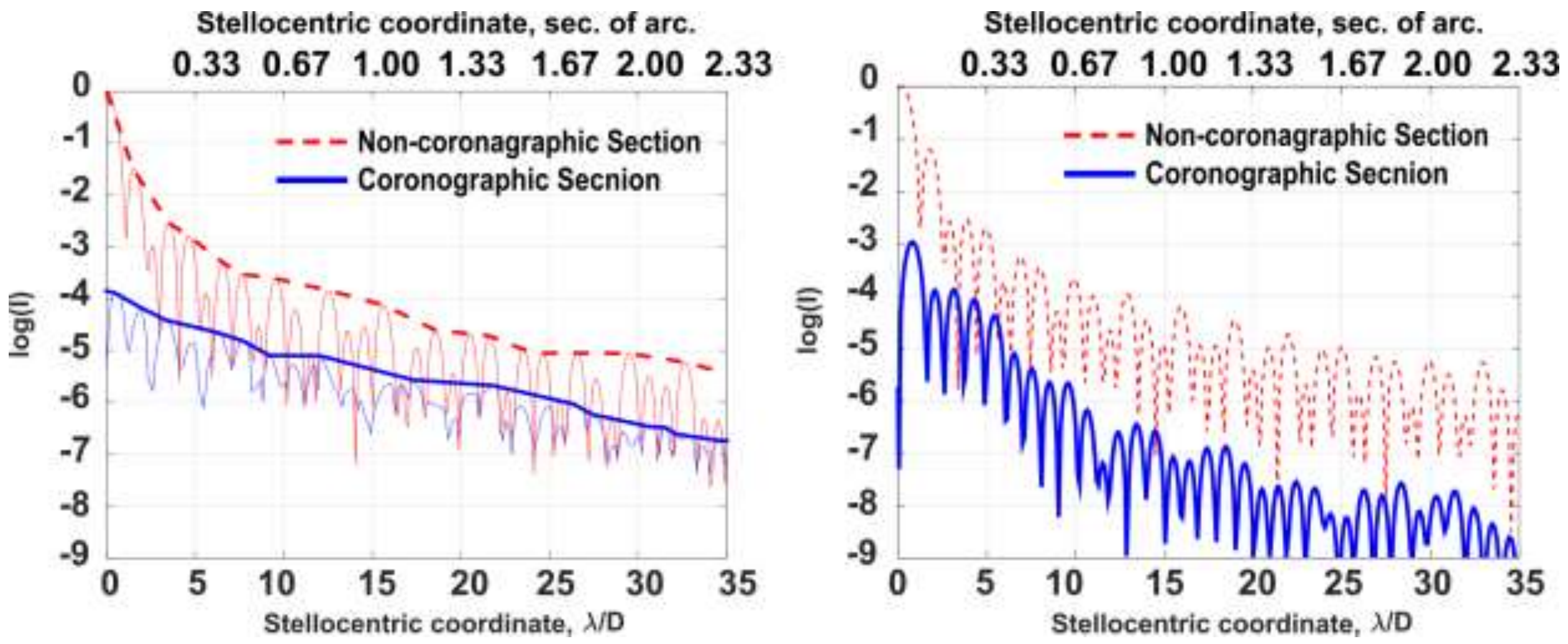

(c) (d)

**Fig. 7** Diffraction effects from spider arms.
(a) – Spider effects not compensated for in coronagraphic images in the plane conjugated to spider arms (left panel), and in the imaging plane (right panel). (b) – Spider effects compensated. A mask to filter out the diffraction effects from spider arms (Opaque mask is shown in transparent green color) (left panel) and (on right panel) image plane plot in the secondary focus with the spider diffraction being mostly compensated (focal mask size is shown by a circle). (c) – PSF without spider diffraction compensation. (d) – PSF with the compensation of spider effects.

## *4.5 Optical Scheme of Interfero-Coronagraph SCEDI*

To clarify the proposed WSO-UV coronagraph SCEDI we show its principal scheme in Fig. 8.

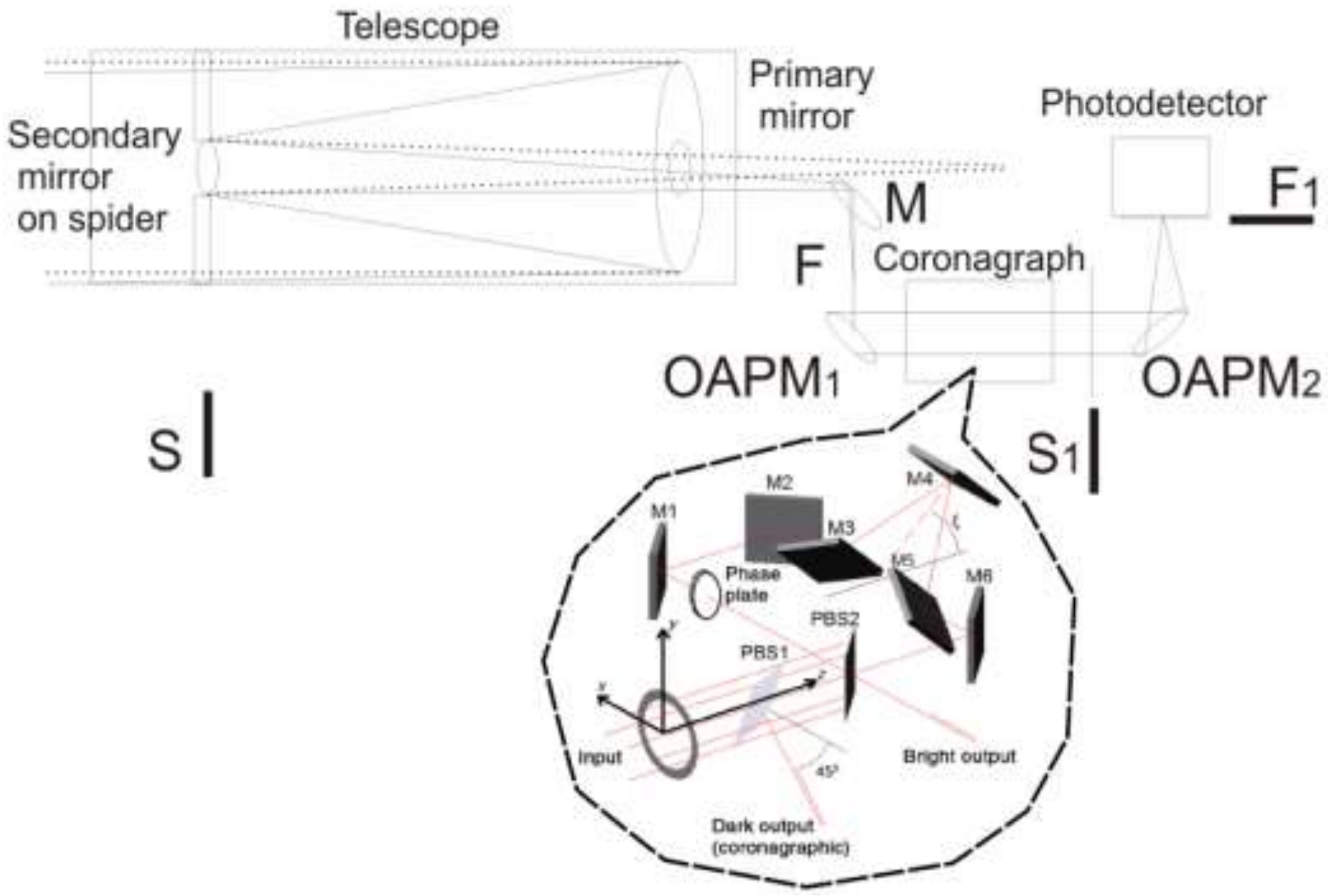

**Fig. 8** SCEDI set-up to insert the interfero coronagraph (PBS polarizing beamsplitter, M – mirror) after the WSO-UV telescope. Assigned optical planes: S – spider "plane", F – primary focus, S1 – plane, optically conjugated to S, F1 – secondary focus.

The optical scheme of interfero-coronagraph SCEDI implements the working principle of the common-path single aperture nulling interferometer described in [11, 12]. This optical scheme reflects the principle of an interfero-coronagraph with variable rotational shear. For simplification in flight, we plan to fix the rotational shear angle at some reasonable value between 5º and 10º. The image of the star and its vicinity is formed by the primary focus of the telescope. This image is collimated and directed to the interferometer. The light from the star is suppressed to reduce destructive interference with antiphase. The light from the two copies of the companion is not suppressed because they are geometrically separated. More precisely, companion PSFs become misaligned, because they are superimposed with the radial shear.

Fig. 8 shows the location of masks needed to cancel out the diffraction from spider arms and to reduce pointing error. The light collected from the WSO-UV telescope in a non-central area (in Pos. 9, Fig. 11) is deflected by a spherical diagonal mirror M to the coronagraph compartment. It is collimated into a parallel beam by an off-axis parabolic mirror OAPM1 and is then transmitted through the coronagraph. In the plane denoted by S1, which is optically conjugated to spider plane S, we have the spider arms sharply re-imaged (see, Fig. 7(b)). Here, we insert the opaque mask having the form shown in Fig. 7(c). It blocks images from two copies of the cross-form spider, rotated to $\pm 2\xi$. The coronagraphic image plane with compensated spider diffraction is formed by the second OAPM2 in the secondary focus F1 coinciding with the photo-detector plane. In this plane (or in its proximity), we add the focal mask to obscure the central spot of Ø $6 \cdot \lambda/D$ at $\lambda$=550 nm to compensate for the residual pointing error.

*4.6 SCEDI Instrument Performance Records*

Herewith we briefly resume previous partial analyses of the coronagraphic instrument to conclude the bottom-line performance estimation.

SCEDI coronagraph will work achromatically in the visible spectral range of 420–700 nm.

Its throughput at 10 λ/D (≈0.67'' @ λ=550 nm) tends to 18%, accounting for the planet's PSF maxima. The coronagraph will have a zero throughput anywhere closer than 3 λ/D (≈0.17''), because of the mask. The intermediate throughput at 5 λ/D (≈0.33'') will be around 6%.

The raw coronagraphic contrast at 10 λ/D (≈0.67'') is between $10^{-5}$–$10^{-6}$ corresponding to pessimistic and optimistic estimations. This considers a possible figure error change and a degraded PSD exponent lower -11/3. At 5 λ/D (≈0.33'') the raw contrast is $10^{-4.5}$–$10^{-5}$, at 20 λ/D (≈1.3'') the raw contrast is $10^{-6.5}$–$10^{-7.5}$. Post-processing can enhance the coronagraphic contrast by two orders or even better by considering both the conditions for space-based optics and a stronger signal from a companion or exoplanet achievable in an achromatic wavelength range.

Therefore, the SCEDI coronagraph w/o adaptive optics can be in range of the declared science goal discussed in Sect. 3.1.

## 5 UV Spectrograph UVSPEX

High sensitivity (photon counting) is good for an M-type star faint in UV. A spectral resolution of 0.5 nm is sufficient to separate major emission lines of exospheric atoms, in particular to separate the O I line from other strong spectral lines. The spectral resolution will be achievable by spectrometers in the main WUVS block [17] (shown in Fig. 1); however, it is difficult to measure the weak flux from M-type stars without a photon-counting detector. To realize exoplanet transit observations in oxygen spectral lines with the desired accuracy, we equip the WSO-UV telescope with the UVSPEX spectrograph.

The main engineering requirements for the UV spectrograph are the following. The spectral resolution is better 0.5 nm. The spectral range is to exceed wavelengths from 115 nm to 135 nm to detect at least H Lyman alpha 121.6 nm to O I 130 nm. The throughput is better than 0.3%, accounting for more than four terrestrial exoplanets distanced at 5 pc.

To achieve these requirements, a simple spectrograph design is proposed, containing a slit, a concave (toroidal) grating as a dispersion element, and an imaging photo-detector. This optical concept is conventional and has been used in other space missions for UV spectroscopy, e.g., Extreme ultraviolet spectrosCope for ExosphEric Dynamics (EXCEED) onboard Hisaki [43] and Probing of Herman Exosphere by Ultraviolet Spectroscopy (PHEBUS) onboard Mercury Planetary Orbiter (MPO) for the BepiColombo mission [44]. Among them, two flight tested components are the grating (CLASP mission [ 45 , 46 ]), and the photo-detector (BepiColombo/PHEBUS-FUV [47, 48], PROCYON/LAICA [13]).

Fig. 9 shows the UVSPEX principal optical scheme. The spectrometer slit is aligned at the primary focus of the telescope from the off-axial sub-FoV, at Pos 10 (see Fig. 11). Slit width is 0.2 mm, corresponding to 5 arc-sec. The concave grating is a laminar type with groove density of 2,400 grooves per mm. It has a toroidal shape with curvature radii of 266.4 mm in horizontal direction and of 253.0 mm in vertical direction. The effective area has nearly Ø 25 mm and the focal length is ~250 mm. The surface is coated by Al + $MgF_2$ to increase the grating reflectance, and the diffraction efficiency of ~29% can be achieved based on its past heritage [46]. To achieve higher diffraction efficiency, a blazed grating is under development as an option. As a detector, we use a conventional photon counting detector with microchannel plates (MCPs) because it produces lower dark noise counts (typically ~1 count/sec/$cm^2$) than other sensors such as CCDs. This type of detector has been used in various space applications [49, 50]. The detector consists of an $MgF_2$ input window, a CsI photocathode, a 5-stage MCP (V- and Z- stacks), and a resistive anode encoder (RAE). Although recently other types of MCP detectors have become available [51], we selected a conventional readout system with an RAE as the baseline design to simplify the system and reduce necessary resources. The core parts of the detector are sealed under vacuum to avoid degradation of the CsI photocathode. The target spectral range of the spectrometer is above 115 nm, which is the cutoff wavelength of the $MgF_2$ window, so we can use a sealed tube detector. The spatial resolution is 80 μm and it is sufficient to achieve the required spectral resolution of 0.5 nm at 130 nm. Quantum efficiency is attainable at 16% at 130 nm. By assuming the reflectivity of the telescope mirrors (Al + $MgF_2$ coating) as 75% at 130 nm, the total throughput and effective area of the spectrometer are 2.0% and 444 $cm^2$, respectively. We calculated the signal-to-noise ratio (SNR) for the UVSPEX system taking into account the MCP noise of 1 count/sec/$cm^2$, and the result is shown in Fig. 10. In our calculation we assume an exposure time of 50 hours in total.

Because UVSPEX is an optional instrument mainly dedicated to detect the weak flux from M-type stars, it is important to clarify its advantage compared to the main spectrometer WUVS. Thus, we also show the SNR of WSO-UV/WUVS [14, 15] in Fig. 10. It is clear that UVSPEX achieves higher SNR than WUVS for lower stellar flux (darker targets) mainly because UVSPEX uses an MCP detector while WUVS uses a CCD detector. Even for the flux from Proxima Centauri, one of the brightest targets of UVSPEX, which is ~$3 \times 10^{-15}$ erg/s/$cm^2$ at 130 nm, we need to use an MCP detector to achieve sufficient SNR for darker targets. Fig. 10 shows the SNR of the flux at 130 nm from the star, and we need sufficient SNR to detect the planets, e.g., the necessary SNR is approximately 100 for Proxima Centauri. To detect the extended oxygen exosphere of Earth-like exoplanets which have a transit depth of ~25% (Sect. 3.2 and Fig. 3), the observations of target stars by UVSPEX and WUVS require SNRs of ~10 and ~3, respectively. Because WUVS has higher spectral resolution than UVSPEX, the transit depth of Earth-like exoplanets in WUVS spectral data is deeper (~75%) and thus lower SNR is required. We have to note that it is difficult in this observation to detect the Venus-like or Mars-like exoplanets with such SNR because of their low transit depth (~0.2%). This defines the detection limits of both instruments as $2.5 \times 10^{-17}$ erg/s/$cm^2$ for UVSPEX and $9.0 \times 10^{-17}$ erg/s/$cm^2$ for WUVS. If we assume a star with the same flux as Proxima Centauri at a longer distance, the distance of detection limits for UVPSEX is 14 pc, almost twice that for WUVS (7.4 pc). Thus, we conclude that UVSPEX has higher capability than WUVS to detect expanded oxygen exospheres of Earth-like planets orbiting around M-dwarfs.

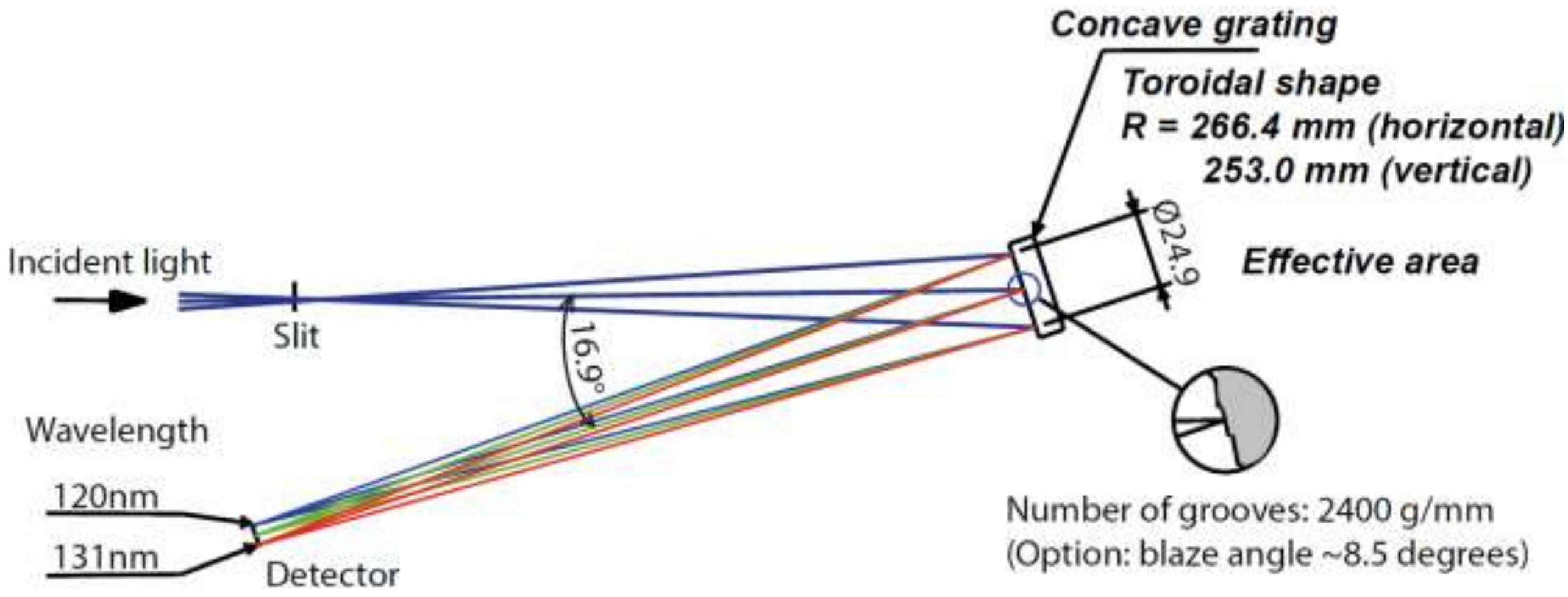

**Fig. 9** Optical scheme of the spectrograph UVSPEX.

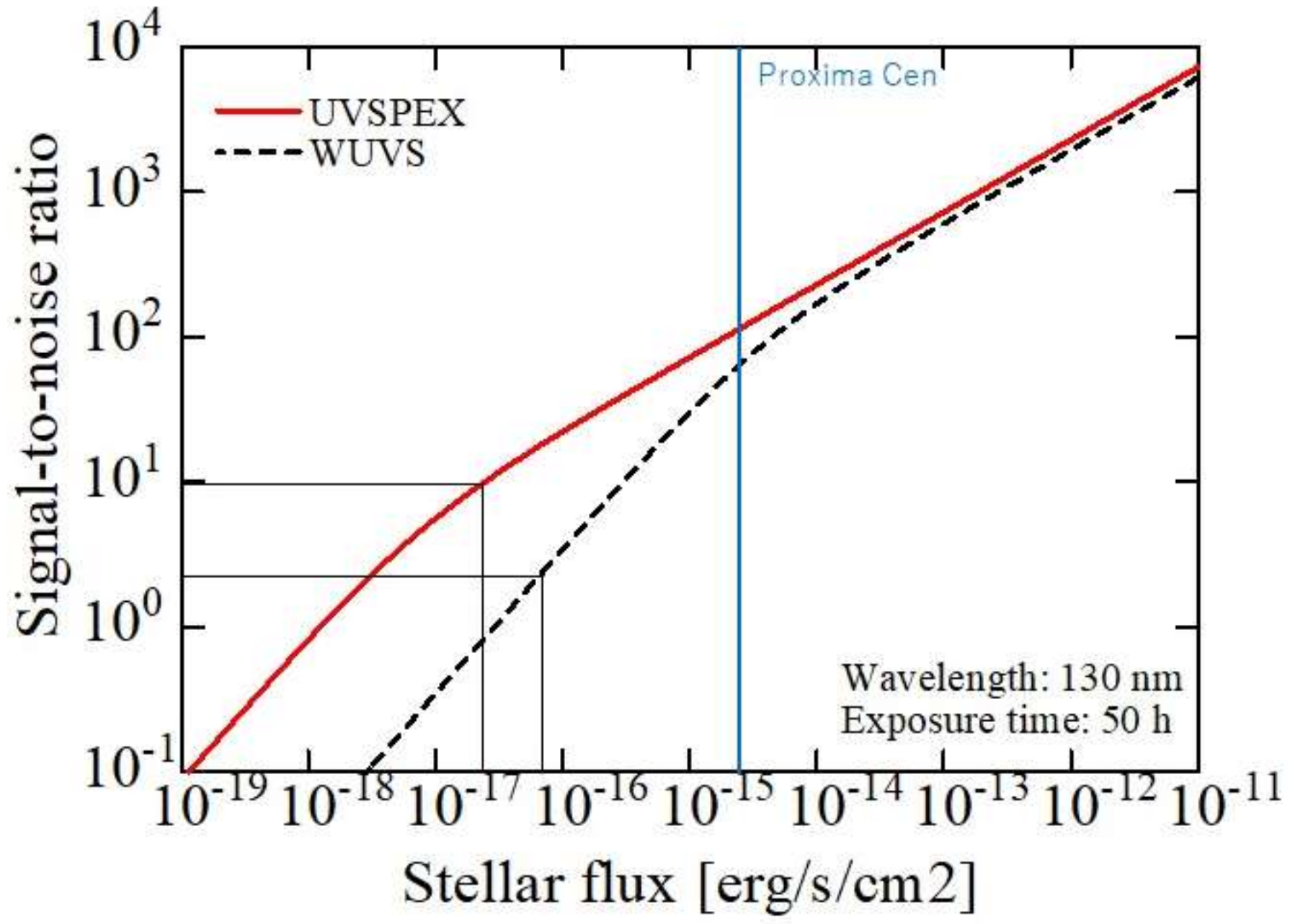

**Fig. 10** Signal-to-noise ratio (SNR) of the UVSPEX and WUVS instruments. Horizontal axis shows the photon flux from the target star at 130 nm. The vertical axis shows the SNR for target stars. Sufficient SNR is required to detect the exosphere of an exoplanet, depending on its transit depth.

## 6 Miscellaneous

To simplify both SCEDI and UVSPEX interfaces, and to minimize power consumption, control units, data downlink, costs, volume, etc., we plan to use a common electronic block serving the two instruments. It will control two similar imaging detectors with different photo-cathodes. One detector will be spectrally optimized for the UVSPEX spectrograph and the other for the SCEDI coronagraph. We will be able to operate the two instruments simultaneously.

The instruments can be technically accommodated in the WSO-UV, and their implementation is pending JAXA approval.

## Conclusion

We have described an imaging stellar coronagraph, *SCEDI*, and a coronal UV spectrometer, *UVSPEX*. Both instruments are aimed to study exoplanetary science onboard the 1.7 m WSO-UV orbital telescope. The WSO-UV orbital telescope is under construction by the Federal Russian Space Program. The scientific goals of the two proposed instruments have been preliminarily defined. Main specifications of the new instruments are discussed within the engineering solutions.

Aiming to enhance UV astronomy, the WSO-UV orbital telescope will have primary and secondary mirrors working in visible wavelengths (within reflectivity >90%) and allow good quality diffraction-limited imaging. The goal of the stellar coronagraph is to create an achromatic high-contrast coronagraphic image of a circumstellar vicinity in the visible band. The coronagraph will be an in-flight demonstrator, and will attempt to photometrically characterize several known giant planets. In the optimistic case, it will image exoplanets and circumstellar discs of several tens of nearby stars. The coronagraph will not implement adaptive optics because of the strict mission simplification requirements. The coronagraphic effect becomes possible at the expense of coronagraph spatial resolution (IWA). The proposed principle is based on a common-path achromatic rotational shear interfero-coronagraph. With the angle of the rotational shear fixed between 5 and 10 degrees, the CPARC demonstrates reduced sensitivity to low-order aberrations from a telescope's optics. The raw coronagraphic contrast can reach a $10^7$ peak-to-peak value at 7–10 $\lambda/D$ stellocentric distances. The coronagraph thus becomes less sensitive to the pointing error. The placement of the coronagraph in an off-axis field of view is proposed.

The goal of the UV-Spectrograph is to study exosphere transit photometric curves of the oxygen emission line to differentiate between different types of rocky planets. FUV irradiation intensity, especially H-Lyman alpha, is also important for photochemistry in the lower atmosphere. H-Lyman alpha (122 nm) and O I (130 nm) emission lines fall within the spectral range of UVSPEX, and if we assume a star with the same flux as Proxima Centauri at a longer distance, the distance of the detection limit for UVSPEX is 14 pc.

To achieve these requirements, a simple spectrograph design was proposed, containing a slit, a concave (toroidal) grating as a dispersion element, and an imaging photo-detector. UVSPEX's spectral range must exceed wavelengths from 115 nm to 135 nm to detect at least the H Lyman alpha 121.6 nm to OI 130 nm lines. The throughput of 2.0% is better than 0.3%, which can account for more than four terrestrial exoplanets around nearby stars.

At the current development stage, not all specifications are known for the entire mission, so the final performance will likely require modifications, but a baseline has been presented for adding these two options to the WSO-UV telescope.

*Acknowledgments*

We acknowledge the support from the Government of Russian Federation and Ministry of Education and Science of Russian Federation (grant N14.W03.31.0017) and the Russian Science Foundation (grant 18-19-00452). This work was partially supported by JSPS KAKENHI 16H04059, 17KK0097, and the MEXT-Supported Program for the Strategic Research Foundation at Private Universities, 2014-2018 (S1411024).

*Appendix 1* *WSO-UV Fields of View. Correction of Optical Aberrations in a Non-Axial FoV of a Richie-Chrétien Telescope for the Coronagraph Instrument SCEDI*

The WSO-UV 1.7 m Ritchey–Chrétien telescope collects light for several instruments and spacecraft systems having imaging cameras and detectors. The WSO-UV telescope is designed to re-distribute several sub-fields of view between scientific instruments and systems. Sub-fields of view are schematically shown in Fig. 11 assuming the plane of primary focus. The sub-FoVs are assigned to (Fig. 11): fine guiding sensors, WUFS spectrometers (correspondingly to UVES, VUVES, and LSS [17]), two field cameras (NUV and FUV imagers). The out-of-axis sub-FoVs e.g., in Pos. 9 and 10 can be used by the proposed instruments SCEDI and UVSPEX.

Observation of an object in each of the FOVs without guides requires specific pointing, and therefore, the observation involves time-sharing.

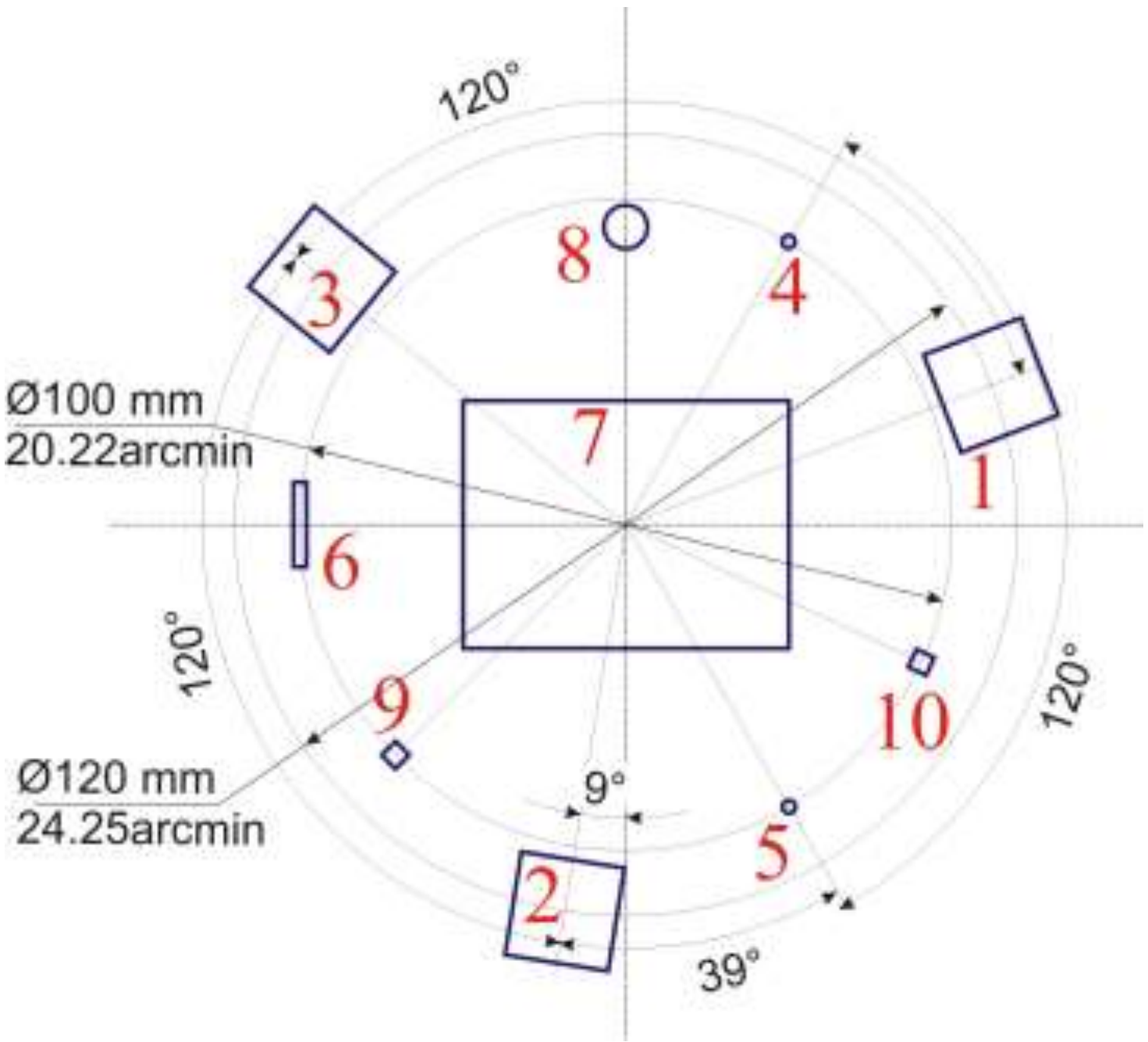

**Fig. 11** Field of view of the WSO-UV 1.7 m telescope in the plane of primary focus, split to sub-FoVs, corresponding to the following instruments and the systems: Pos. 1..3 – fine-guiding sensors (FGSs) w. FoV 3.24 x 3.24 arcmin ≈12.13 arcmin off-axis, Pos. 4 – UV Echelle spectrograph (UVES) w. FoV Ø 1 arcsec ≈10.1 arcmin off-axis, Pos. 5 – vacuum-(far)-UV Echelle spectrograph (VUVES) w. FoV Ø 1 arcsec ≈10.1 arcmin off-axis, Pos. 6 – long-slit spectrometer (LSS) w. FoV 72×1 arcsec 10.11 arcmin off-axis, Pos. 7 – near UV field camera imager (NUV) w. FoV 10×7.7 arcmin on-axis, Pos. 8 – far-UV field camera imager (FUV) w. FoV Ø 1.35 arcmin ≈9 arcmin off-axis, Pos. 9 – stellar coronagraph (SCEDI) w. FoV 20×20 arcsec set ≈10 arcmin off-axis, Pos. 10 – UV coronal spectrograph (UVSPEX) w. FoV 60×2.5 arcsec set ≈10 arcmin off-axis.

We analyzed the geometrical aberrations in the non-axial sub-FoV Pos. 9 (Fig. 11) assigned to the coronagraph instrument. The coronagraph's FoV is required to have aberrations as small as possible, not to exceed the p-v amount, which is given by the residual aberrations from the surface figure error for the axial field (λ/5 at λ=550 nm). The sub-FoV in Pos. 9 for the coronagraph can be small, covering less than 20 x 20 seconds of arc. This sub-FoV is however strongly de-centered from the axis to nearly 10 arc minutes, because the axial sub-FoV (Pos. 7) is occupied by the main field camera, in particular, the near-UV (NUV) field camera imager (Fig. 11).

The geometrical aberrations from a Ritchey–Chrétien telescope of 1.7 m of the WSO-UV design, assuming its ideal surfaces in the off-axis sub-FoV in Pos. 9, are shown in Fig. 12(a)). The waveform map shows the optical aberration of ≈1.4.λ p-v (≈0.31.λ rms) at λ=550 nm. It exceeds the required residual aberrations of λ/5 p-v by more than 7 times.

We found a simple way to correct these optical aberrations by means of a spherical diagonal mirror, denoted by ***M***, being placed before the primary focus. Correction of aberrations was shown to improve the level to λ/7 p-v @ λ=550 nm.  The corrected aberration plot for the sub-FoV in Pos. 9 is shown in Fig. 12(b). Optical evaluations were performed using the Zemax® software.

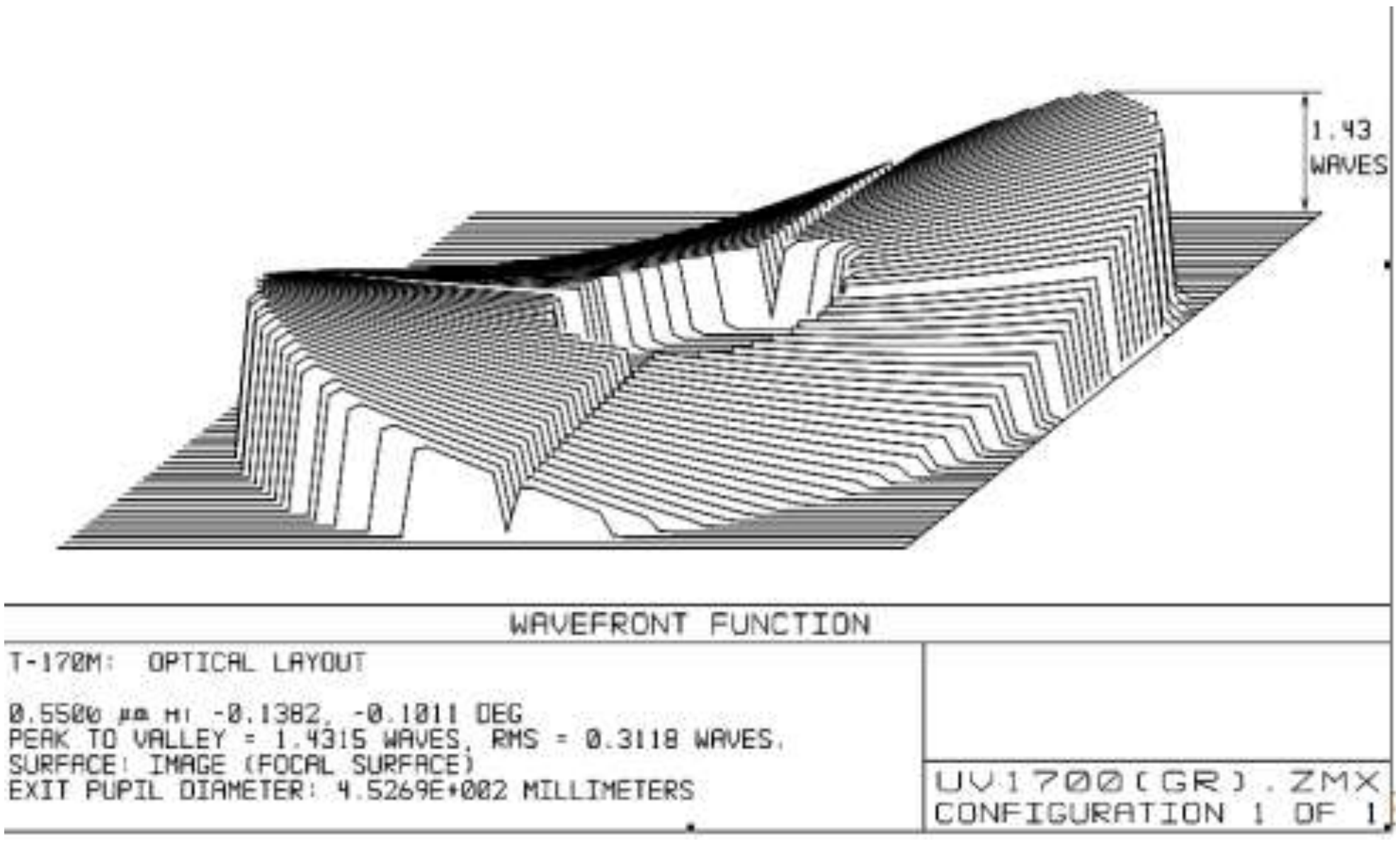
1.43
WAVES
WAVEFRONT FUNCTION
T-170M: OPTICAL LAYOUT
-0.1382, -0.1011 DEG
PEAK TO VALLEY = 1.4315 WAVES, RMS = 0.3118 WAVES.
SURFACE: IMAGE (FOCAL SURFACE)
EXIT PUPIL DIAMETER: 4.5269E+002 MILLIMETERS
UV1700(GR).ZMX
CONFIGURATION 1 OF 1

(a)

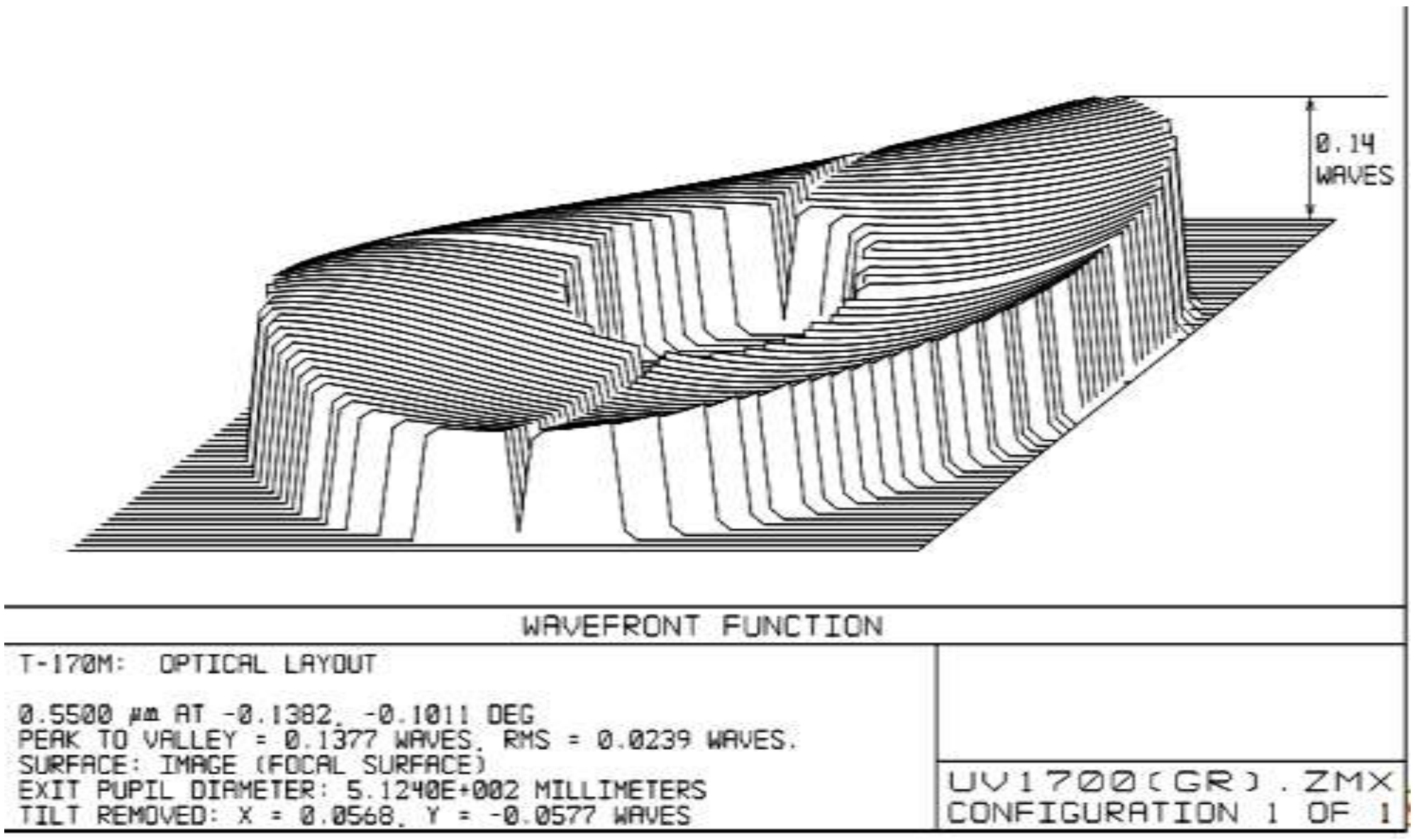
0.14
WAVES
WAVEFRONT FUNCTION
T-170M: OPTICAL LAYOUT
0.5500 µm AT -0.1382, -0.1011 DEG
PEAK TO VALLEY = 0.1377 WAVES, RMS = 0.0239 WAVES.
SURFACE: IMAGE (FOCAL SURFACE)
EXIT PUPIL DIAMETER: 5.1240E+002 MILLIMETERS
TILT REMOVED: X = 0.0568, Y = -0.0577 WAVES
UV1700(GR).ZMX
CONFIGURATION 1 OF 1

(b)

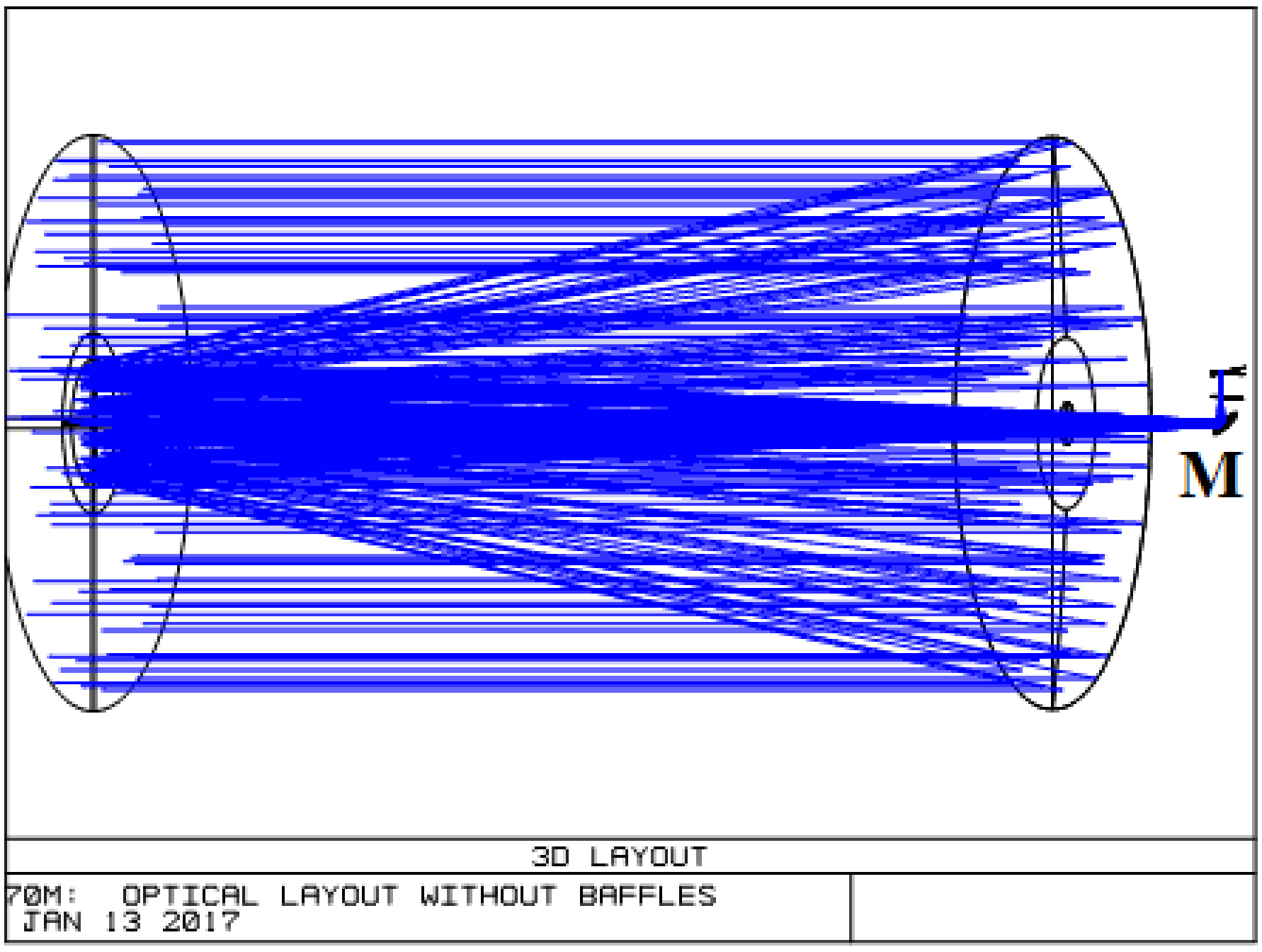

(c)

**Fig. 12** WSO-UV telescope optical scheme: (a) – non-corrected geometrical aberrations waveform having ~1.4.λ p-v (0.31.λ rms) at λ=550 nm for the SCEDI instrument FoV of 20×20 arc seconds, decentered by 10 arc minutes; (b) – corrected geometrical aberrations plot has ~ λ/7 p-v (0.024.λ rms) at λ=550 nm; (c) – WSO-UV telescope optical scheme with a correcting diagonal mirror M shown.

The optical scheme of the Ritchey–Chrétien 1.7 m diameter WSO-UV telescope with a correcting diagonal mirror denoted by "***M***" is shown in Fig. 12(c). Simultaneously, this diagonal mirror ***M*** is also used to deflect the light to the coronagraph instrument compartment.